\documentclass{mn2e}
\usepackage{graphicx,epsfig}
\newcommand{\mbh}{M_{\bullet}}

\begin{document}
\title[Relativistic spectral features from X-ray illuminated spots]
{Relativistic spectral features from X-ray illuminated spots and
the measure of the black hole mass in AGN}

\author[M.~Dov\v{c}iak, S.~Bianchi, M.~Guainazzi, V.~Karas and G.~Matt]
{M.~Dov\v{c}iak,$^{\!1,2}$
 S.~Bianchi,$^{\!3}$ 
 M.~Guainazzi,$^{\!4}$
 V.~Karas$^{1,2}$ and 
 G.~Matt$^{3}$ \\~\\
$^1$~Astronomical Institute, Academy of Sciences of the Czech Republic,
 Bo\v{c}n\'{\i}~II, CZ-140\,31~Prague, Czech Republic\\
$^2$~Charles University, Faculty of Mathematics and Physics, 
 V~Hole\v{s}ovi\v{c}k\'ach~2, CZ-180\,00~Prague, Czech Republic\\
$^3$~Dipartimento di Fisica, Universit\`a degli Studi ``Roma Tre'', 
 Via della Vasca Navale 84, I-00146~Roma, Italy \\
$^4$~XMM Science Operation Center, RSSD-ESA, VILSPA, Apartado 50727,
 E-28080~Madrid, Spain}

\date{Accepted .... Received ...}
\pagerange{\pageref{firstpage}--\pageref{lastpage}} 
\pubyear{2004}
\maketitle
\label{firstpage}

\begin{abstract}
Narrow spectral features in the 5--6~keV range were recently discovered 
in the X-ray spectra of a few active galactic nuclei. We discuss the 
possibility that these features are due to localized spots
which occur on the surface of an accretion disc following
its illumination by flares. We present detailed line profiles 
as a function of orbital phase of the spot and its radial distance
from a central black hole. Comparison of these 
computed profiles with observed features can help to estimate
parameters of the system. In principle this method can provide 
a powerful tool to measure the mass of super-massive black holes 
in active galactic nuclei. By comparing our
simulations with the {\it{}Chandra} and {\it{}XMM--Newton} results, we
show, however, that spectra from present generation X-ray satellites are
not of good enough quality to fully exploit the method and determine
the black hole mass with sufficient accuracy. This task has to be 
deferred to future missions with high throughput and high 
energy resolution, such as {\it{}Constellation--X} and {\it{}Xeus}.

\end{abstract}


\section{Introduction}
Relativistic iron line profiles may provide a powerful tool to 
measure the mass of the black hole in active galactic nuclei (AGNs) 
and Galactic black hole candidates. To this aim,
Stella (1990) proposed to use temporal changes in the line
profile following variations of the illuminating primary source 
(which at that time was assumed to be located on the disc axis
for simplicity). Along the same line of thought, 
Matt \& Perola (1992) proposed to employ, instead, variations of the
integrated line properties such as equivalent width, centroid energy and
line width. These methods are very similar conceptually to the
classical reverberation mapping method, widely and successfully applied
to optical broad lines in AGNs. Sufficiently long
monitoring of the continuum and of the line emission is required, 
as well as large enough signal-to-noise ratio. However, the 
above-mentioned methods have not provided many results yet. Even in the 
best studied case of the Seyfert galaxy MCG--6-30-15, the mass estimate is 
hard to obtain due to the apparent lack of correlation between 
the line and continuum emission (Fabian et al. 2002). It was suggested 
that also these complications are possibly caused by an interplay of 
complex general relativistic effects (Miniutti et al. 2003).
X-ray spectra from high throughput and high energy resolution
detectors should resolve the problem of interpretation of 
observed spectral features. However, before such high quality data are 
available it is desirable to examine existing spectra and attempt to
constrain physical parameters of the models.

A simple, direct and potentially robust way to measure the 
black hole mass would be available if the line emission originates 
at a given radius and azimuth, as expected if the disc
illumination is provided by a localized flare just above the
disc (possibly due to magnetic reconnection), rather than a central
illuminator or an extended corona. If a resulting `hot spot' co-rotates with the
disc and lives for at least a significant part of an orbit, by fitting
the light curve and centroid energy of the line flux, the inclination
angle $\theta_{\rm{}o}$ and the orbit radius could be derived (radius in units of 
the gravitational radius $r_{\rm{}g}$). Further, assuming Keplerian rotation, 
the orbital period is linked with radius in a well-known manner. The equation
for the orbital period then contains the black hole mass $\mbh$ 
explicitely, and so this parameter can be determined, as
discussed later.

Hot spots in AGN accretion discs were popular for a while, following the
finding of apparent periodicity in the X-ray emission of the Seyfert~1
galaxy NGC~6814. They, however, were largely abandoned when this periodicity
was demonstrated to be associated with an AM~Herculis system in the field of view
rather than the AGN itself (Madejski et al. 1993). Periodicities in AGNs 
were subsequently reported in a few sources (Iwasawa et al. 1998; Lee
et al. 2000; Boller et al. 2001). The fact that they were not confirmed in different
observations of the same sources is not
surprising -- quite on the contrary, it would be hard to imagine a hot spot
surviving for several years.

Recently, the discovery of narrow emission features in the X-ray spectra
of several AGNs (Turner et al. 2002, 2004; Guainazzi 2003; Yaqoob et al. 2003) 
has renewed interest in hot spots. There is a tentative
explanation (even if not the only one) for these features,
typically observed in the $5$--$6$~keV energy range, in terms of iron
emission produced in a small range of radii and distorted by 
joint action of Doppler and gravitational shift of photon energy. 
Iron lines would be produced by localized flares 
which illuminate the underlying disc surface, producing the line by
fluorescence. Indeed, the formation of 
magnetic flares on the disc surface is one of the
most promising scenarios for the X-ray emission of AGNs. A particularly
strong flare, or one with a very large anisotropic emission towards the
disc, could give rise to the observed features. Small width of the
observed spectral
features implies that the emitting region must be small, and that
it is seen for only a fraction of the entire orbit 
(either because the flare dies out, or
because emission goes below detectability, see next section). 
If the flares co-rotate with the disc and if they last
for a significant part of the orbit, it may be possible
by observing their flux and energy
variations with phase to determine the orbital parameters, and thence 
$\mbh$.

In section 2 we illustrate the basic properties of the line emitted from an
orbiting, illuminated spot. In section 3 we compare calculations with
the relevant narrow-line features reported in several
AGNs. Section 4 is devoted to the summary and perspectives.

\begin{figure*}
\includegraphics*[width=5.4cm]{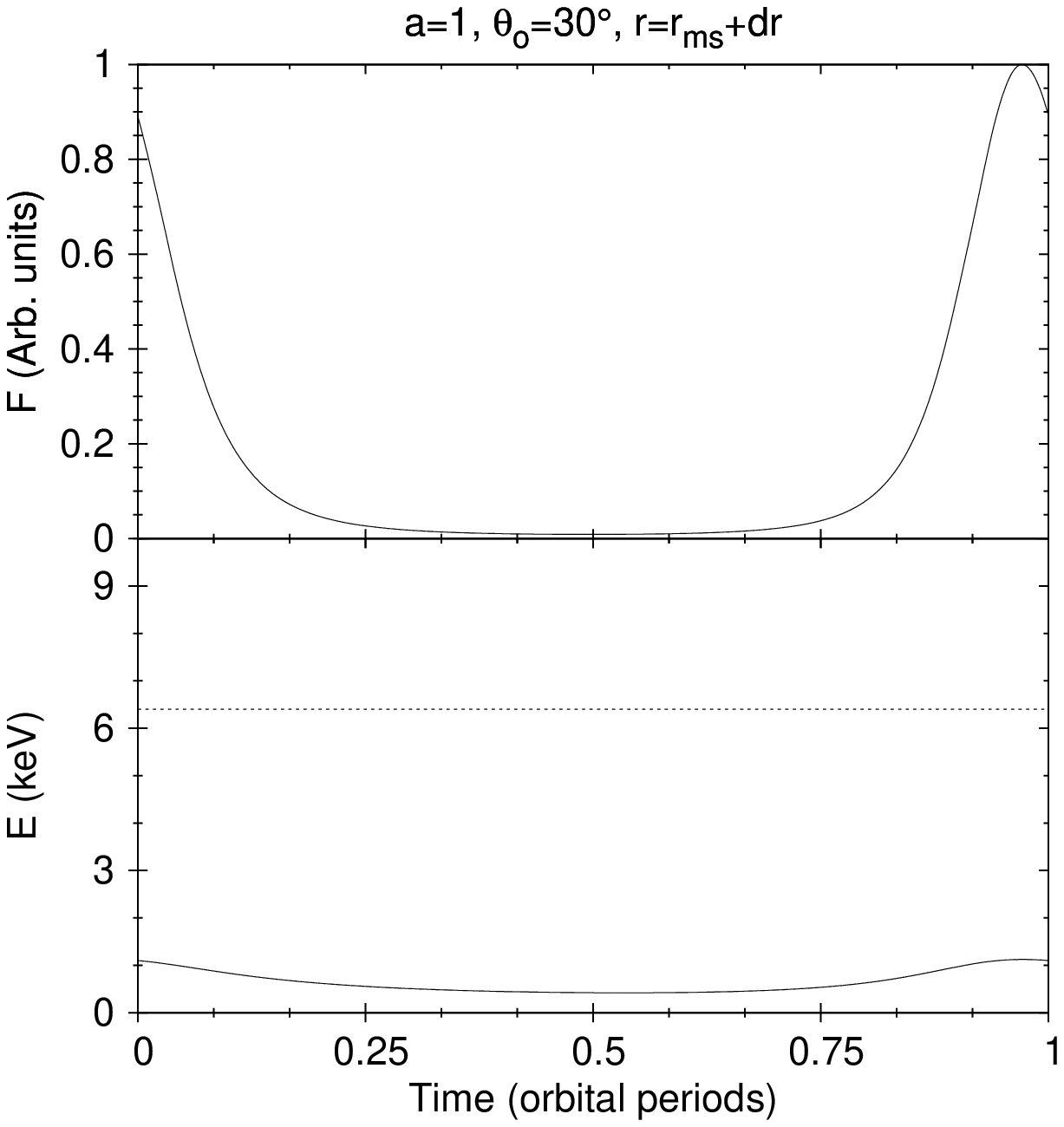}
\hfill
\includegraphics*[width=5.4cm]{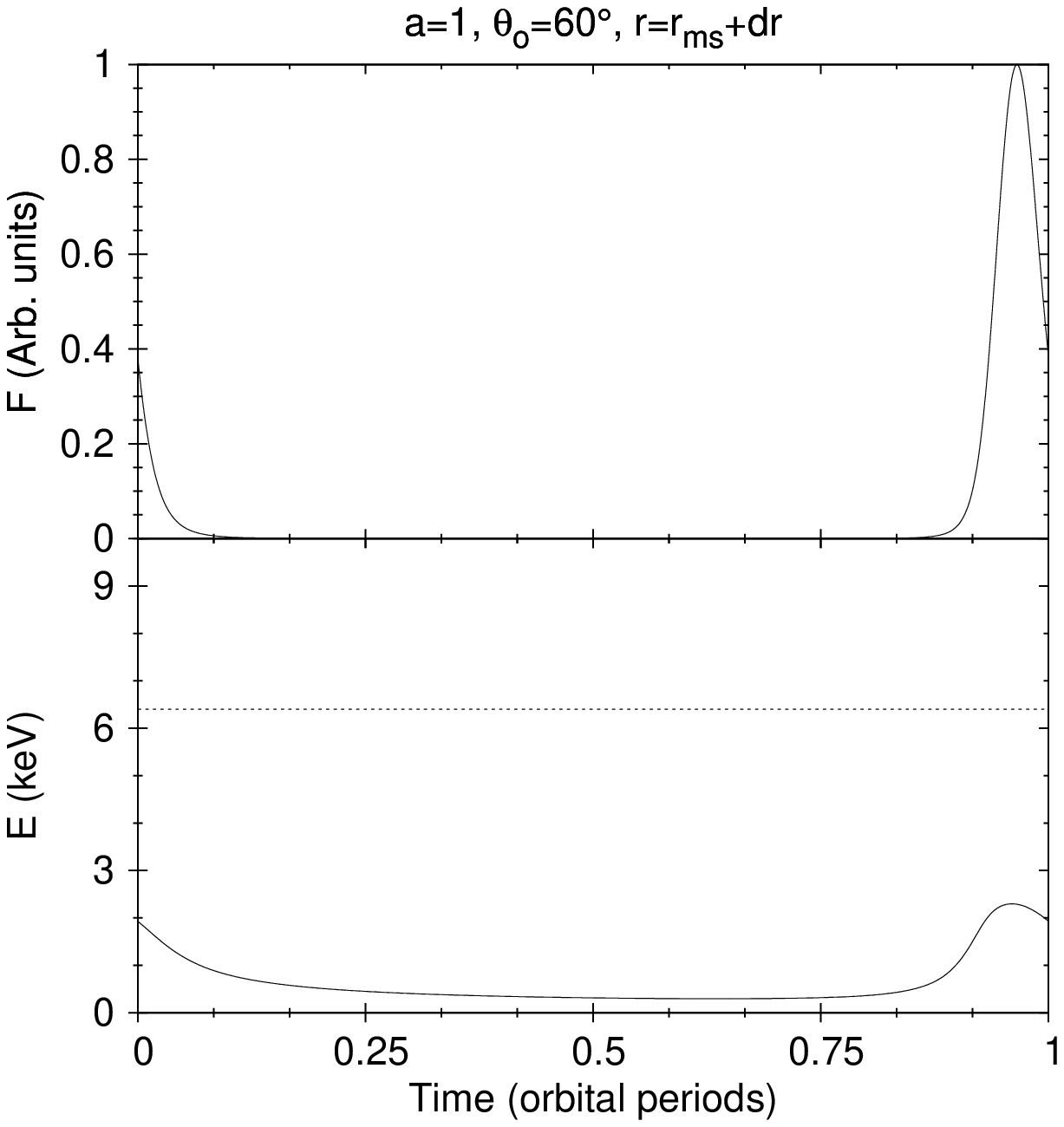}
\hfill
\includegraphics*[width=5.4cm]{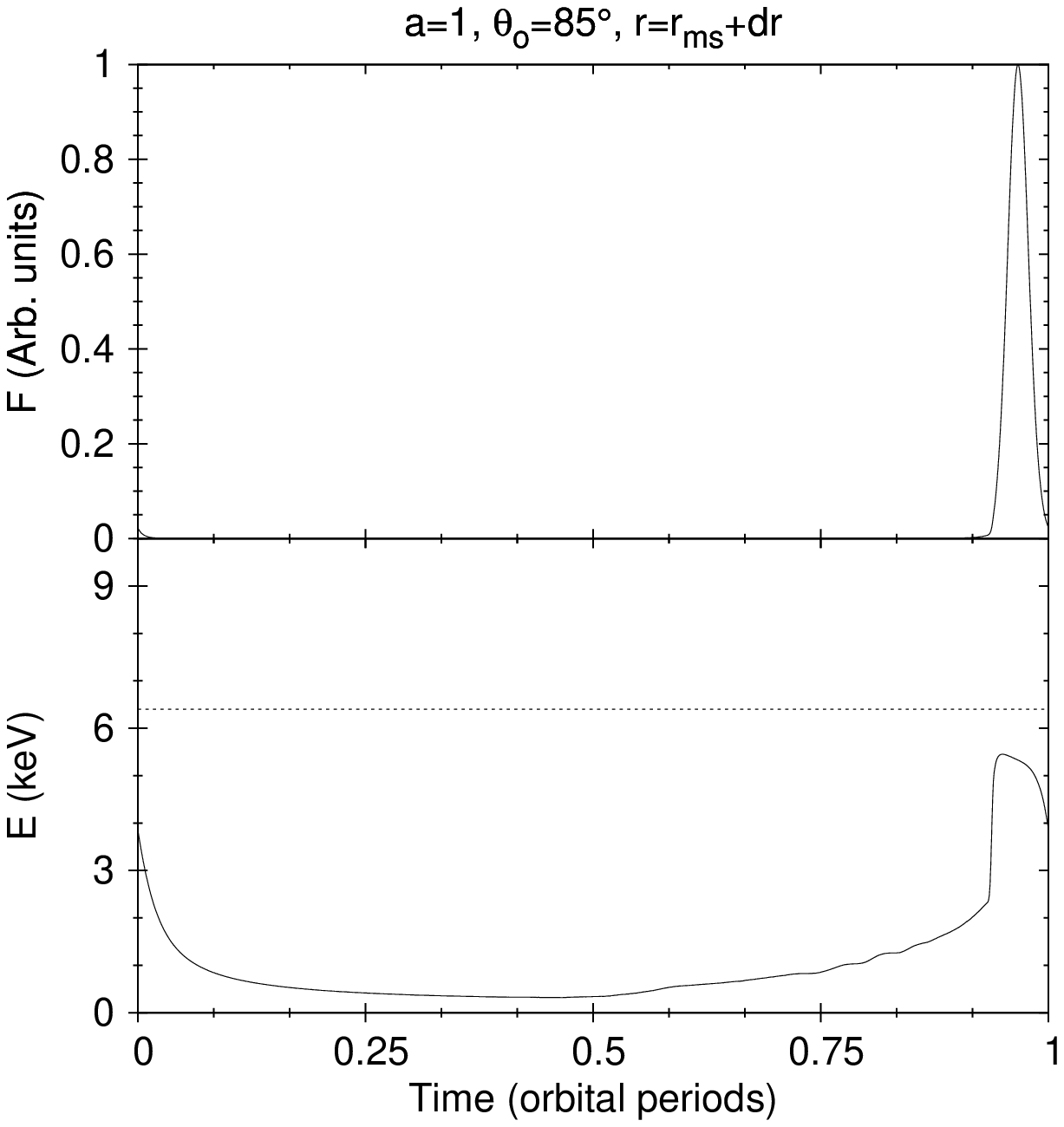}
\vspace*{3mm}\\
\includegraphics*[width=5.4cm]{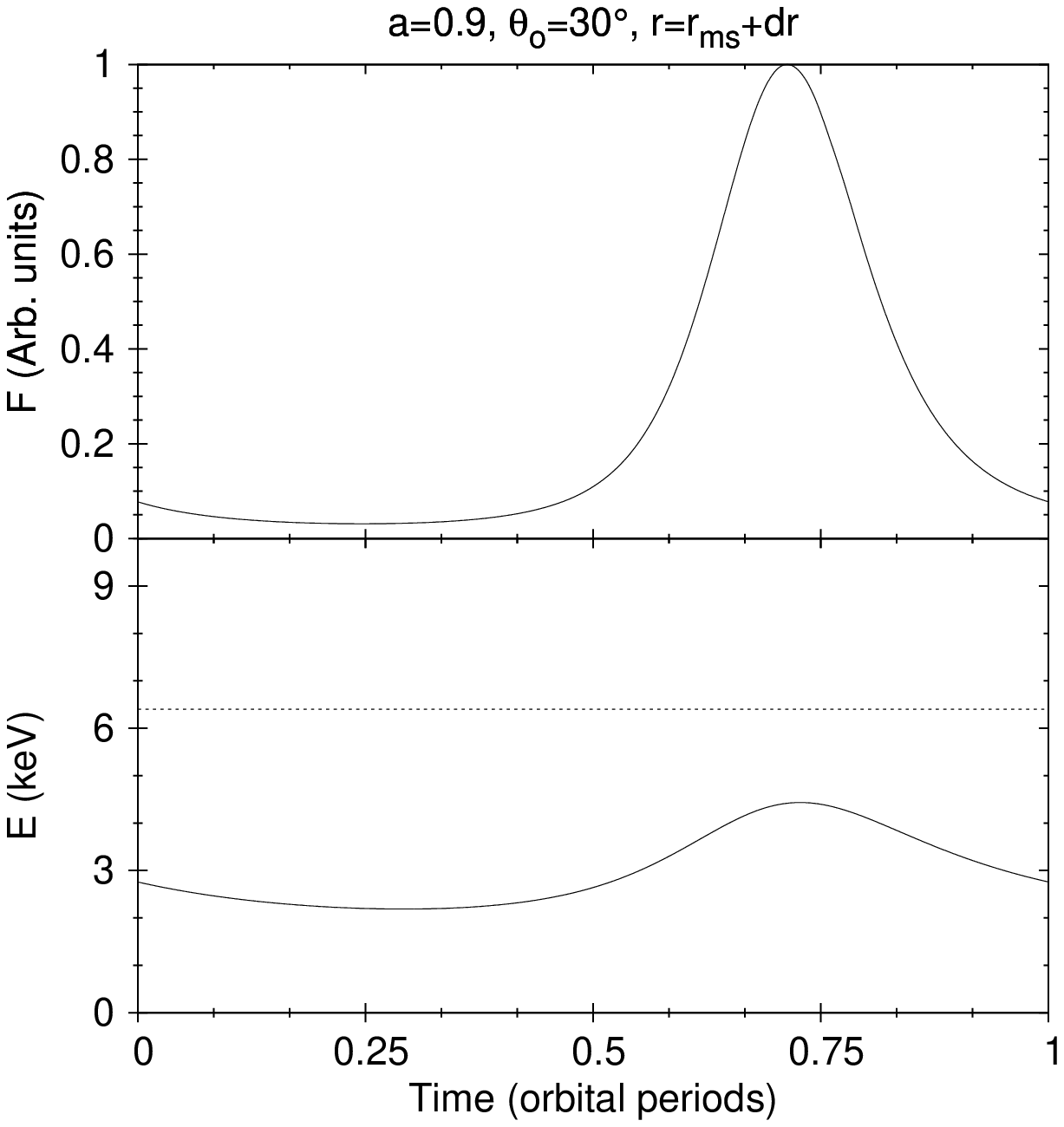}
\hfill
\includegraphics*[width=5.4cm]{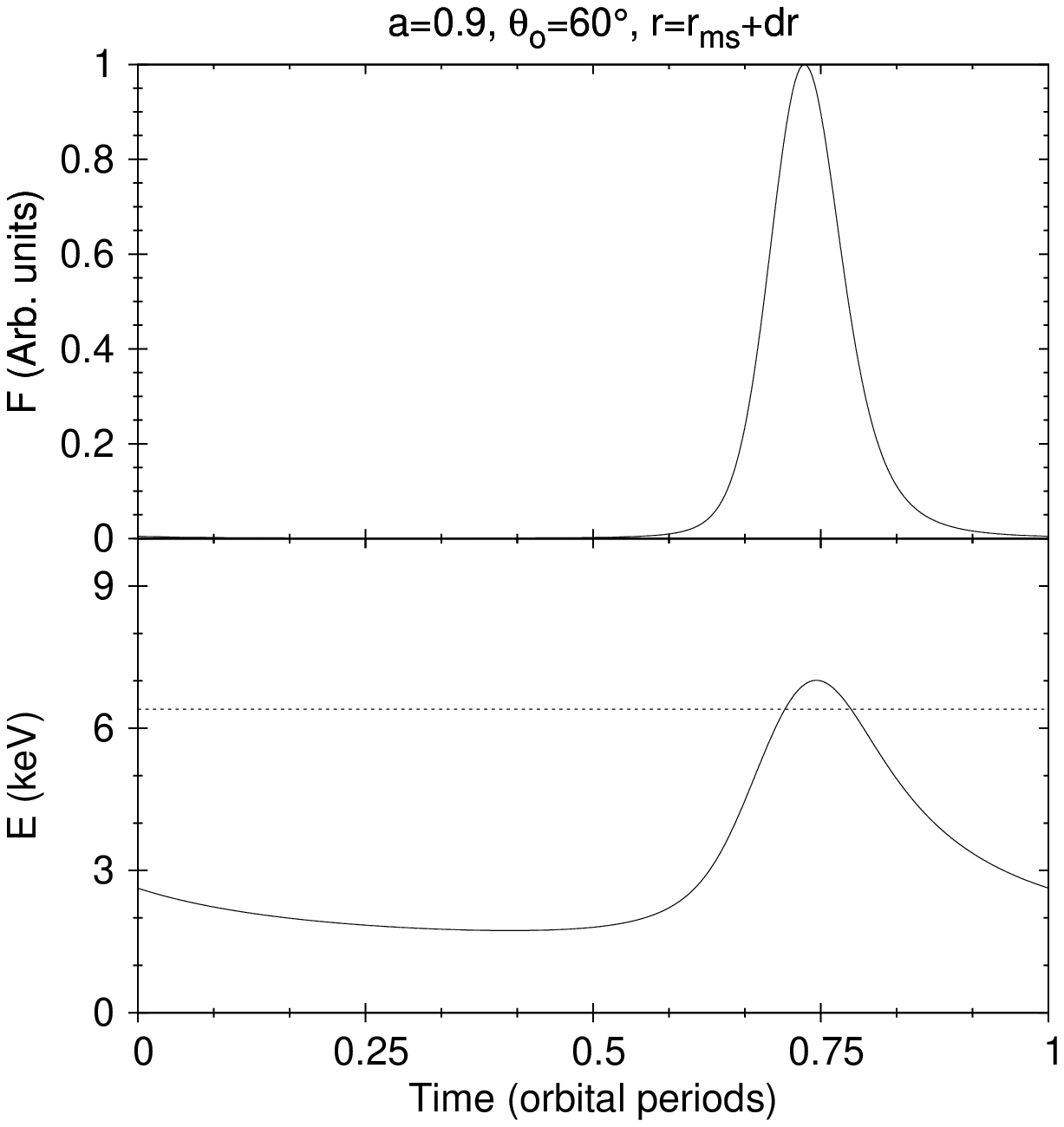}
\hfill
\includegraphics*[width=5.4cm]{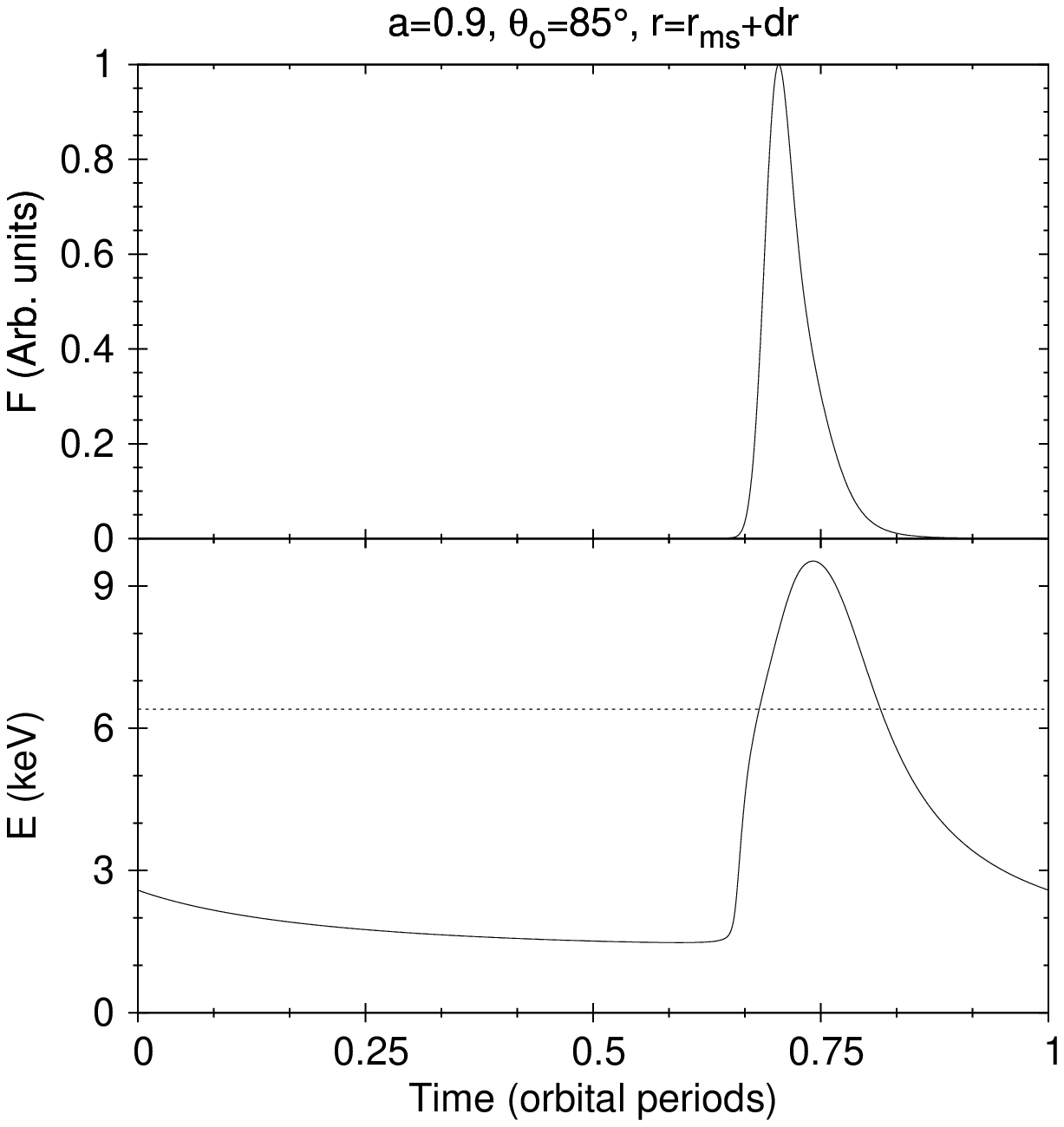}
\vspace*{3mm}\\
\includegraphics*[width=5.4cm]{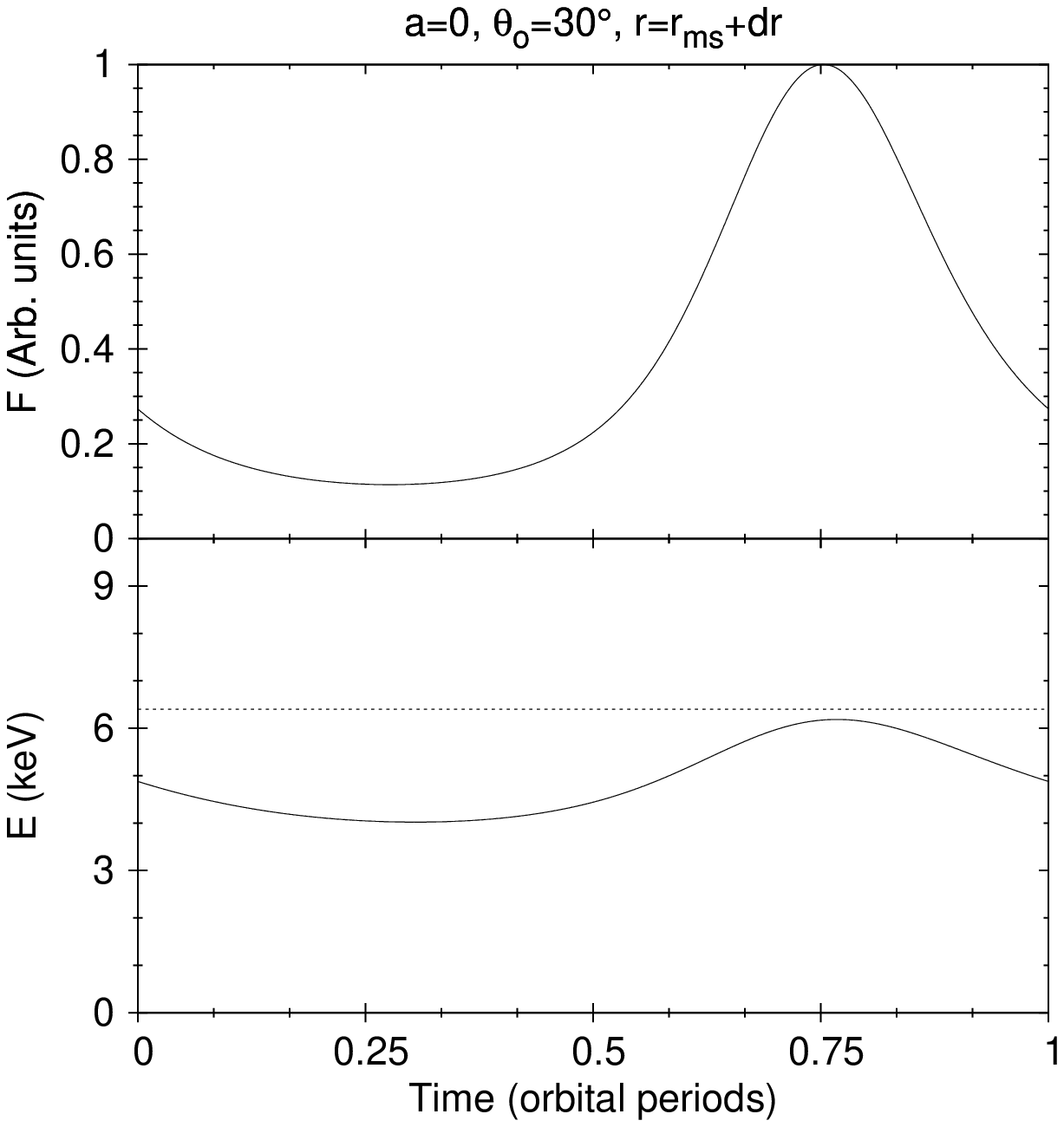}
\hfill
\includegraphics*[width=5.4cm]{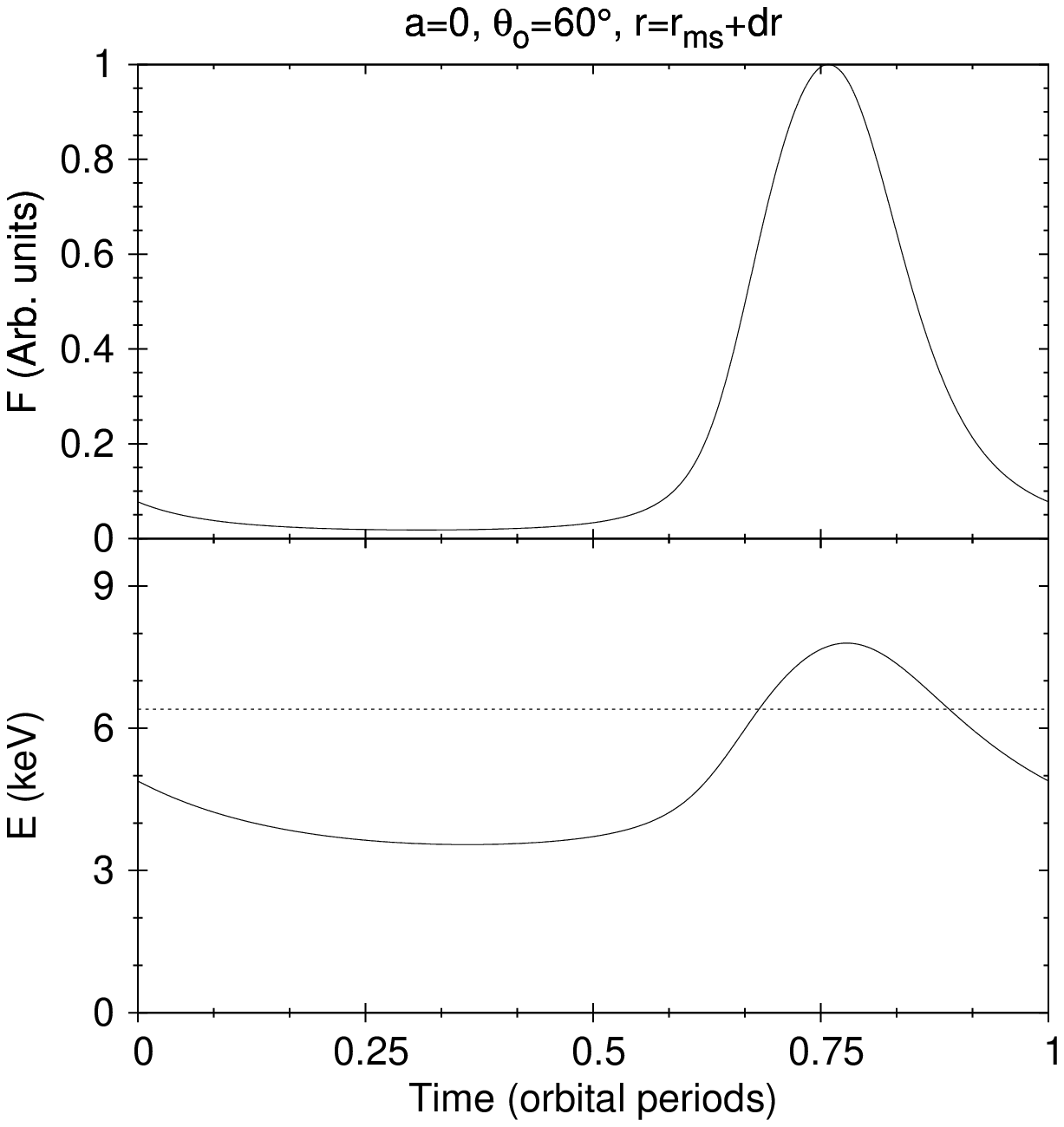}
\hfill
\includegraphics*[width=5.4cm]{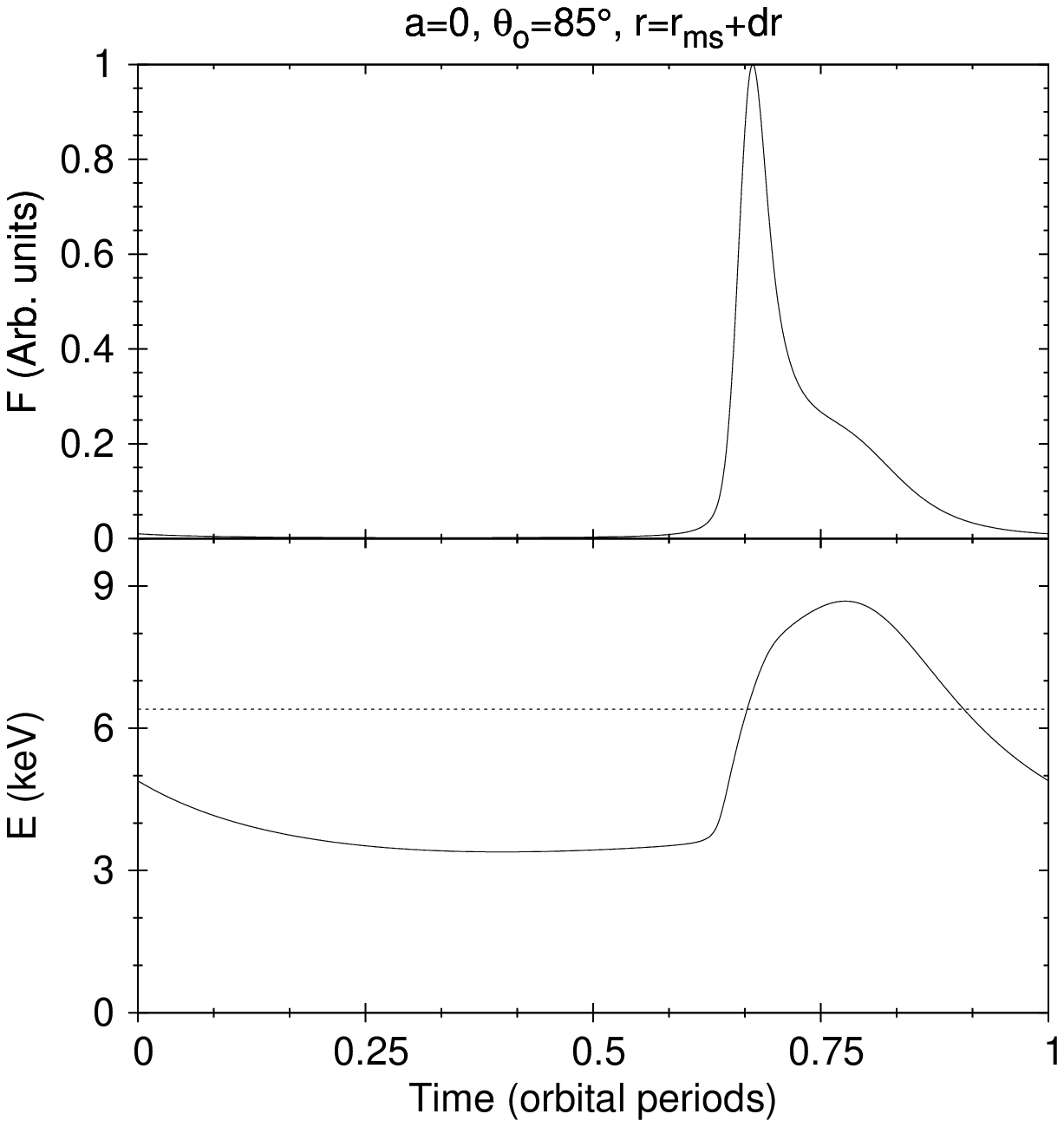}
\caption{Line flux and centroid energy as functions of the 
orbital phase of a spot, for three values of angular momentum 
($a=1$, $0.9$, and $0$) and three inclination angles 
($\theta_{\rm{}o}=30^{\circ}$, $60^{\circ}$, $85^{\circ}$). 
The center of the spot is located at radial distance $r$, which corresponds 
to the last stable orbit $r_{\rm{}ms}(a)$ for that angular momentum
plus a small displacement given by the spot radius, ${\rm{}d}r$. 
The intrinsic energy of the line emission is assumed to be at 
$6.4$~keV (indicated by a dotted line). Prograde rotation is assumed. 
Time is expressed in orbital periods.}
\label{orbits_1}
\end{figure*}

\begin{figure*}
\includegraphics*[width=5.4cm]{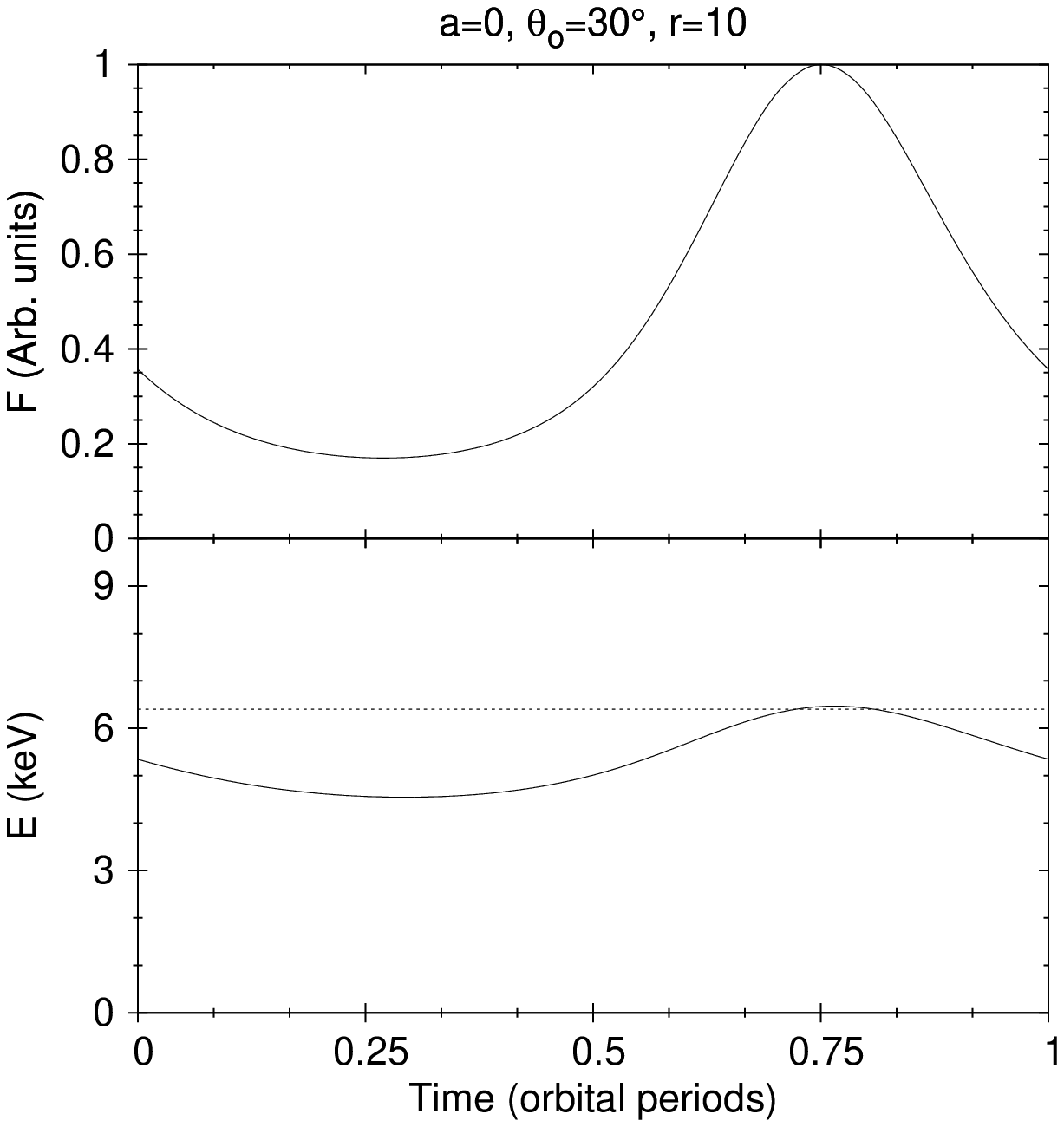}
\hfill
\includegraphics*[width=5.4cm]{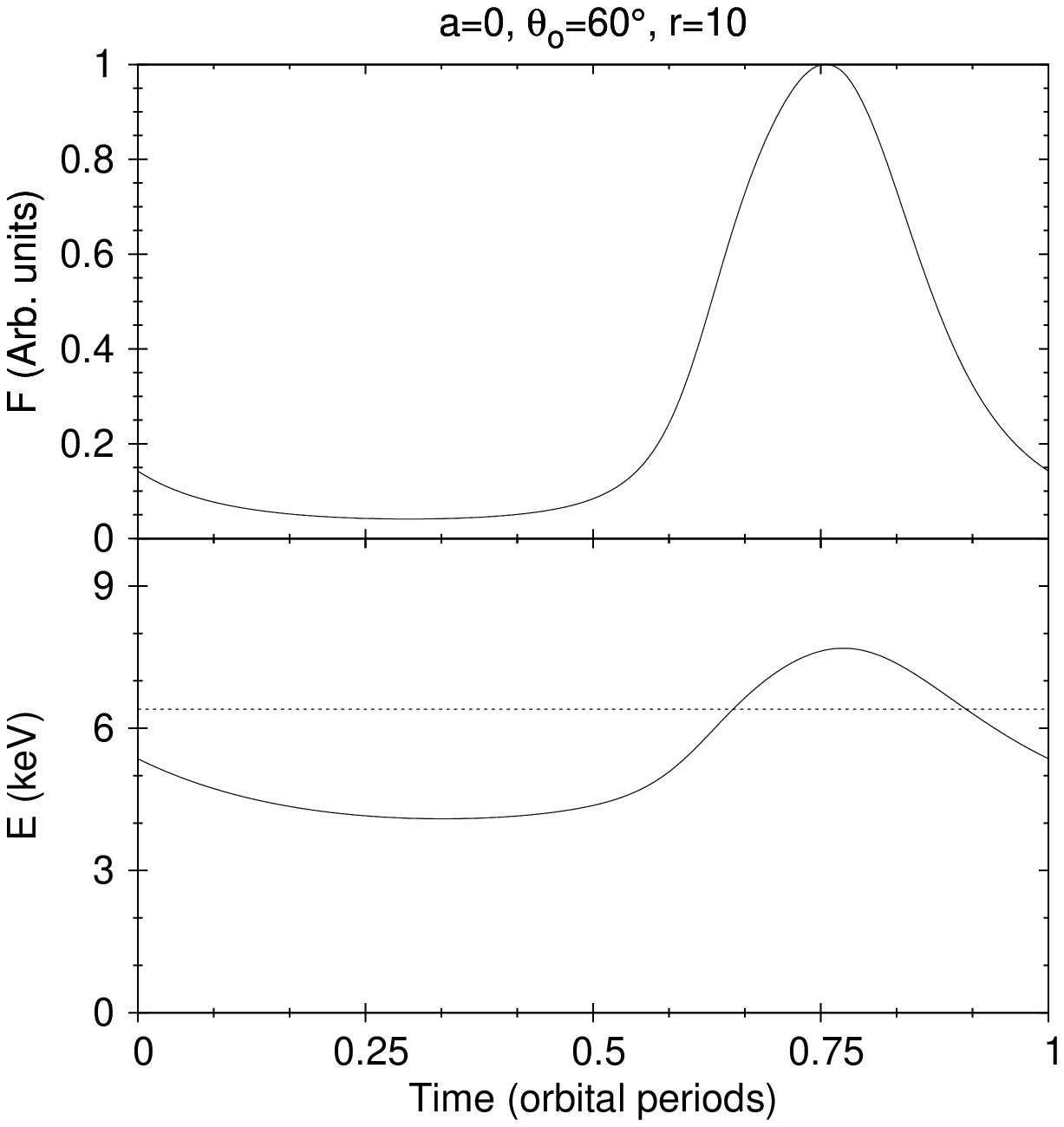}
\hfill
\includegraphics*[width=5.4cm]{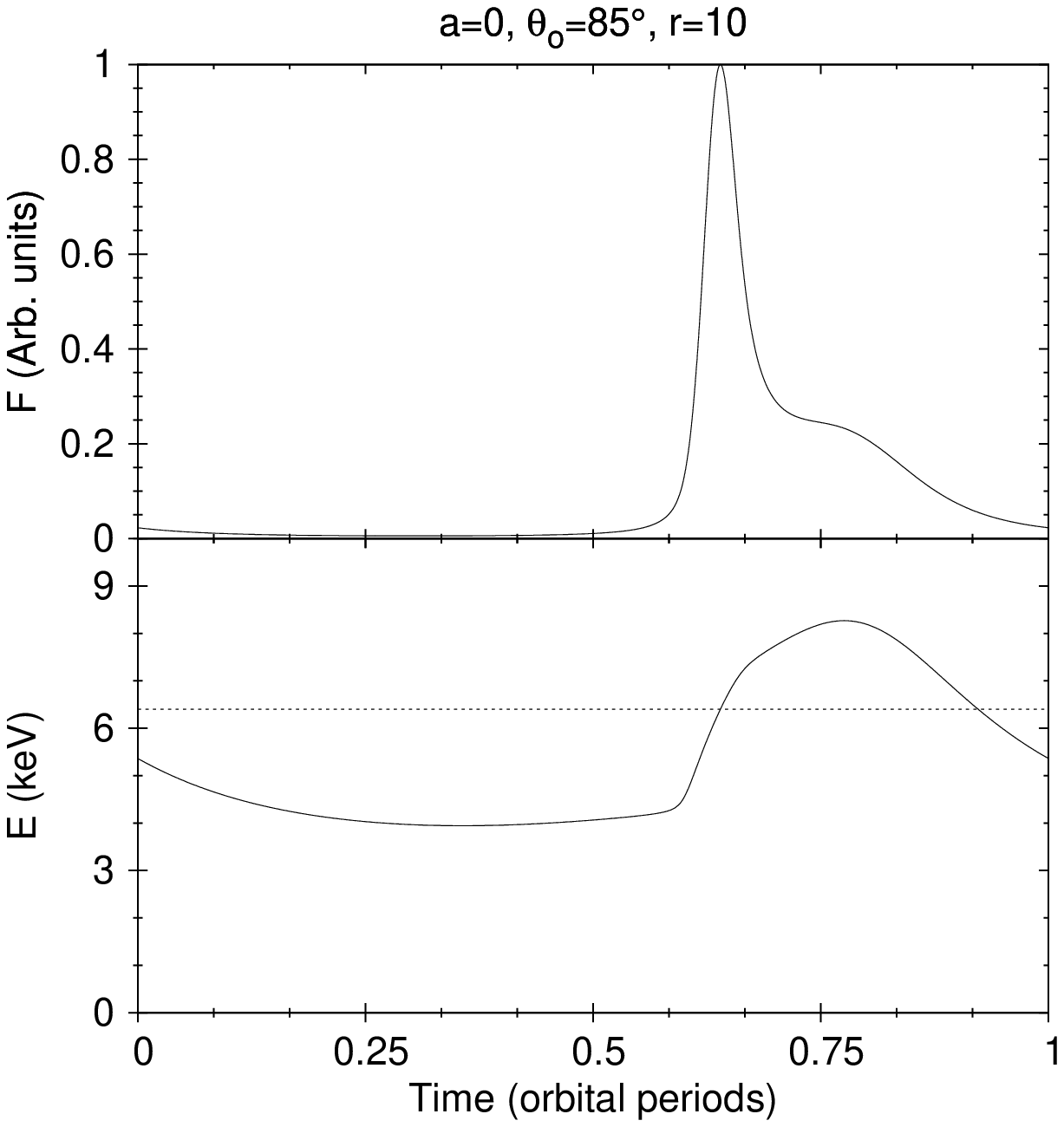}
\vspace*{3mm}\\
\includegraphics*[width=5.4cm]{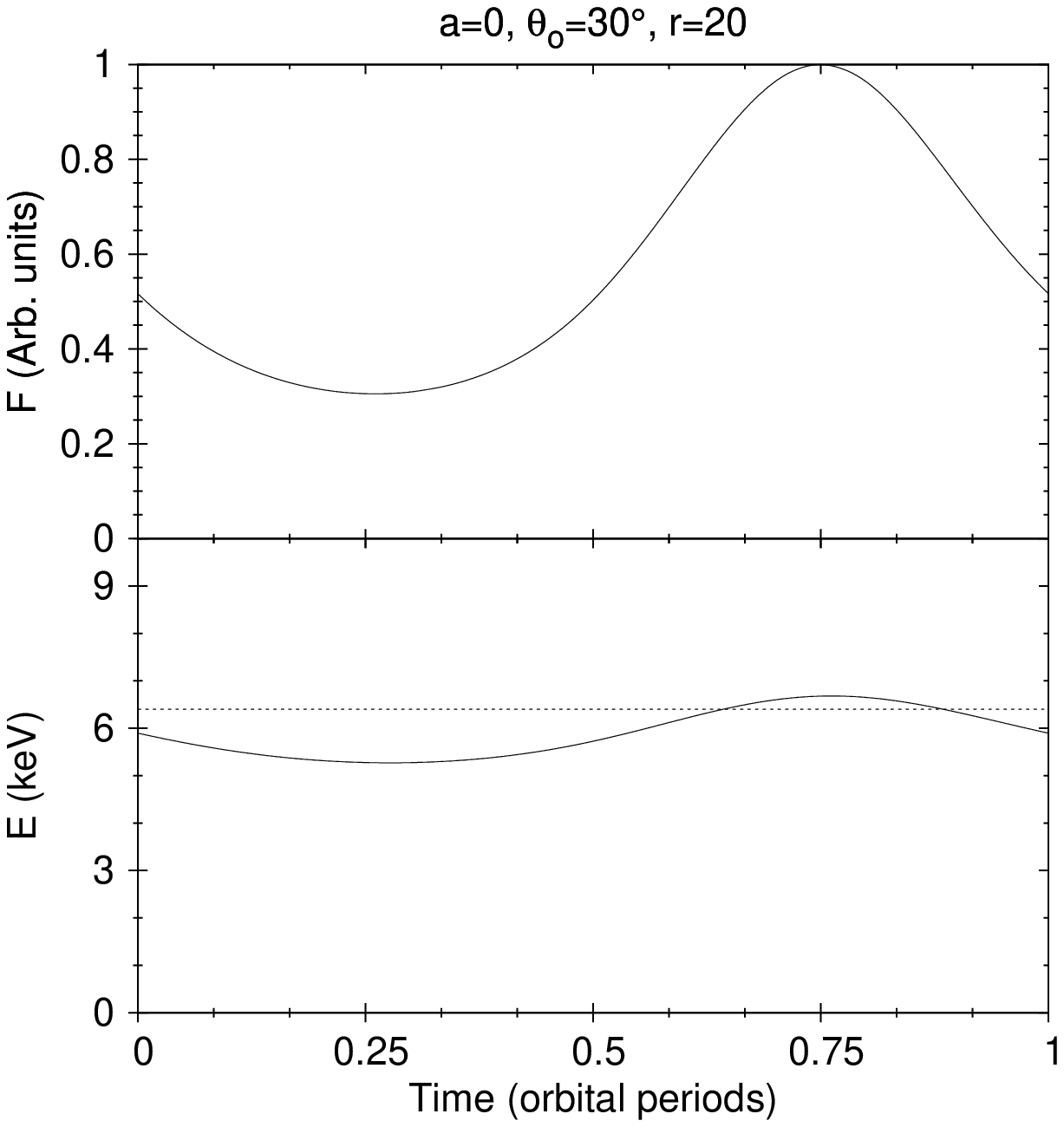}
\hfill
\includegraphics*[width=5.4cm]{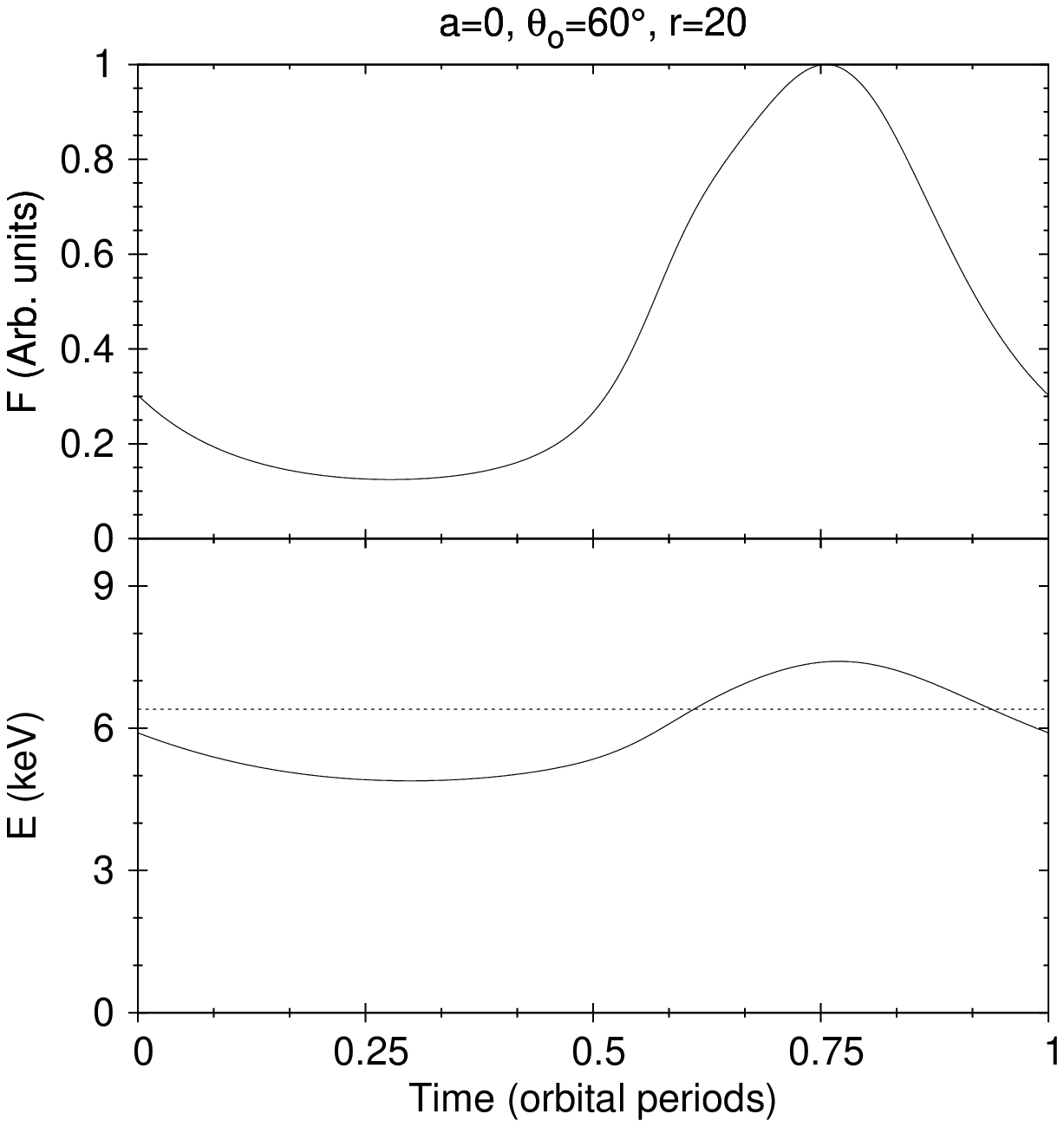}
\hfill
\includegraphics*[width=5.4cm]{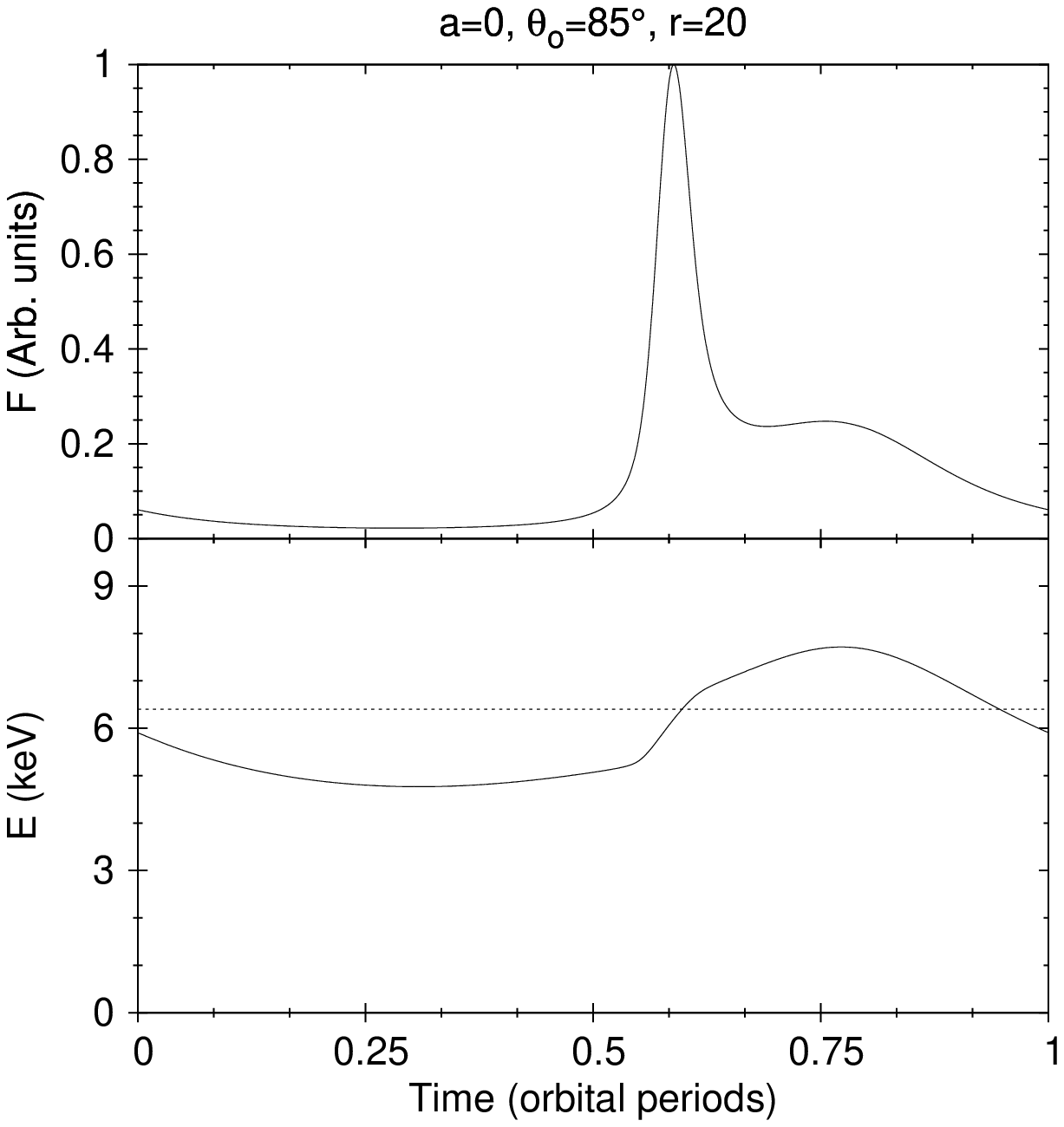}
\vspace*{3mm}\\
\includegraphics*[width=5.4cm]{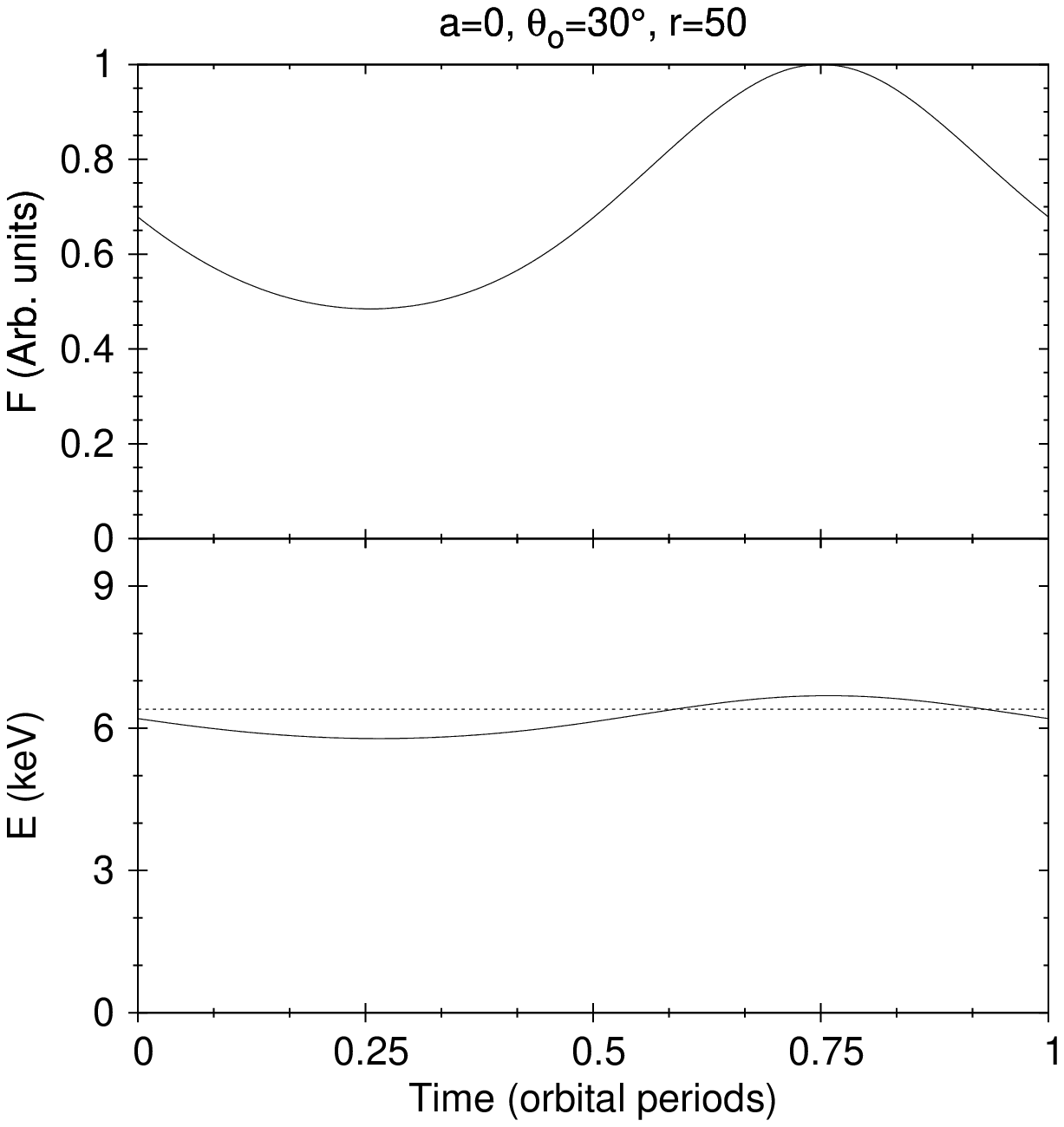}
\hfill
\includegraphics*[width=5.4cm]{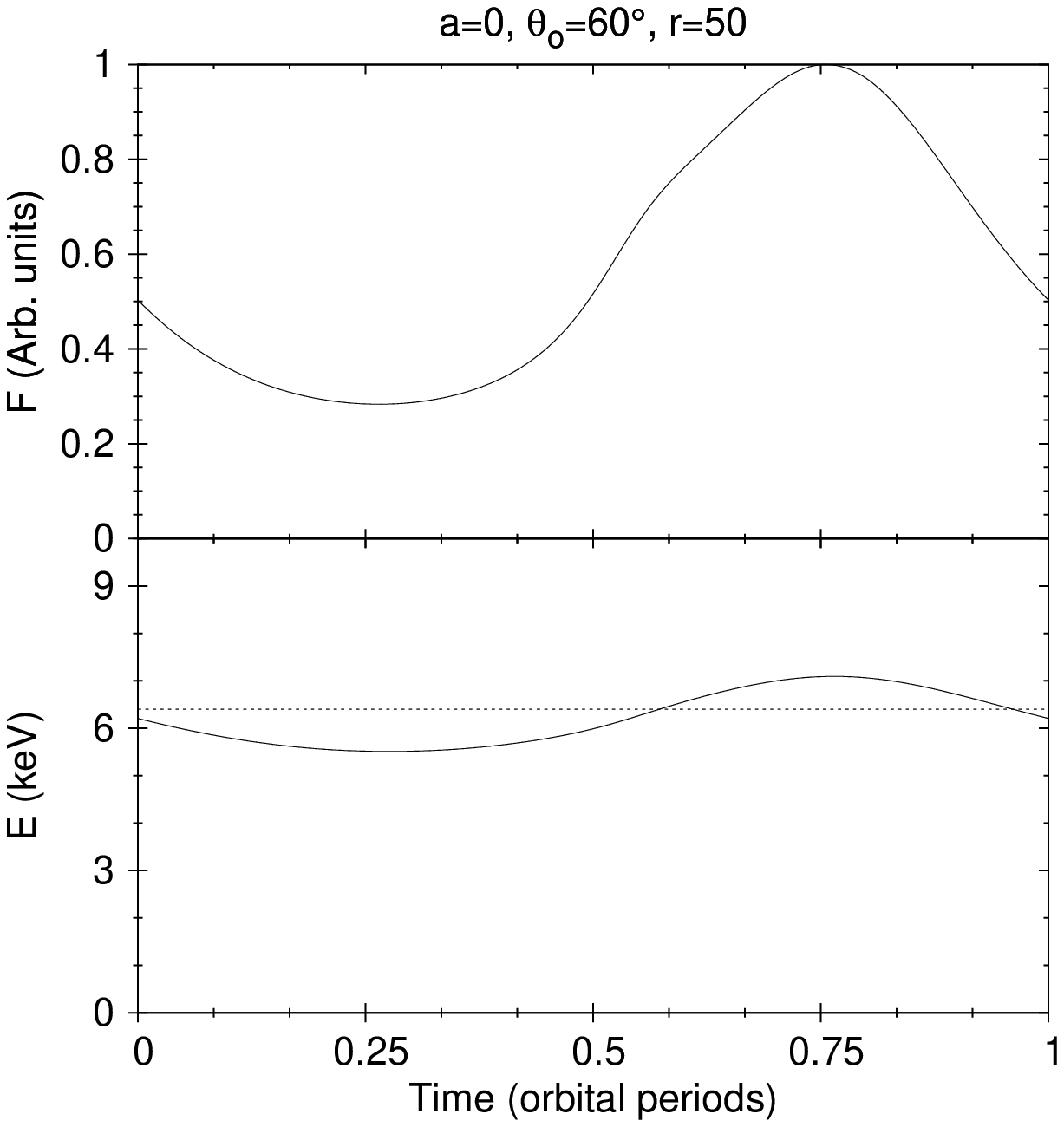}
\hfill
\includegraphics*[width=5.4cm]{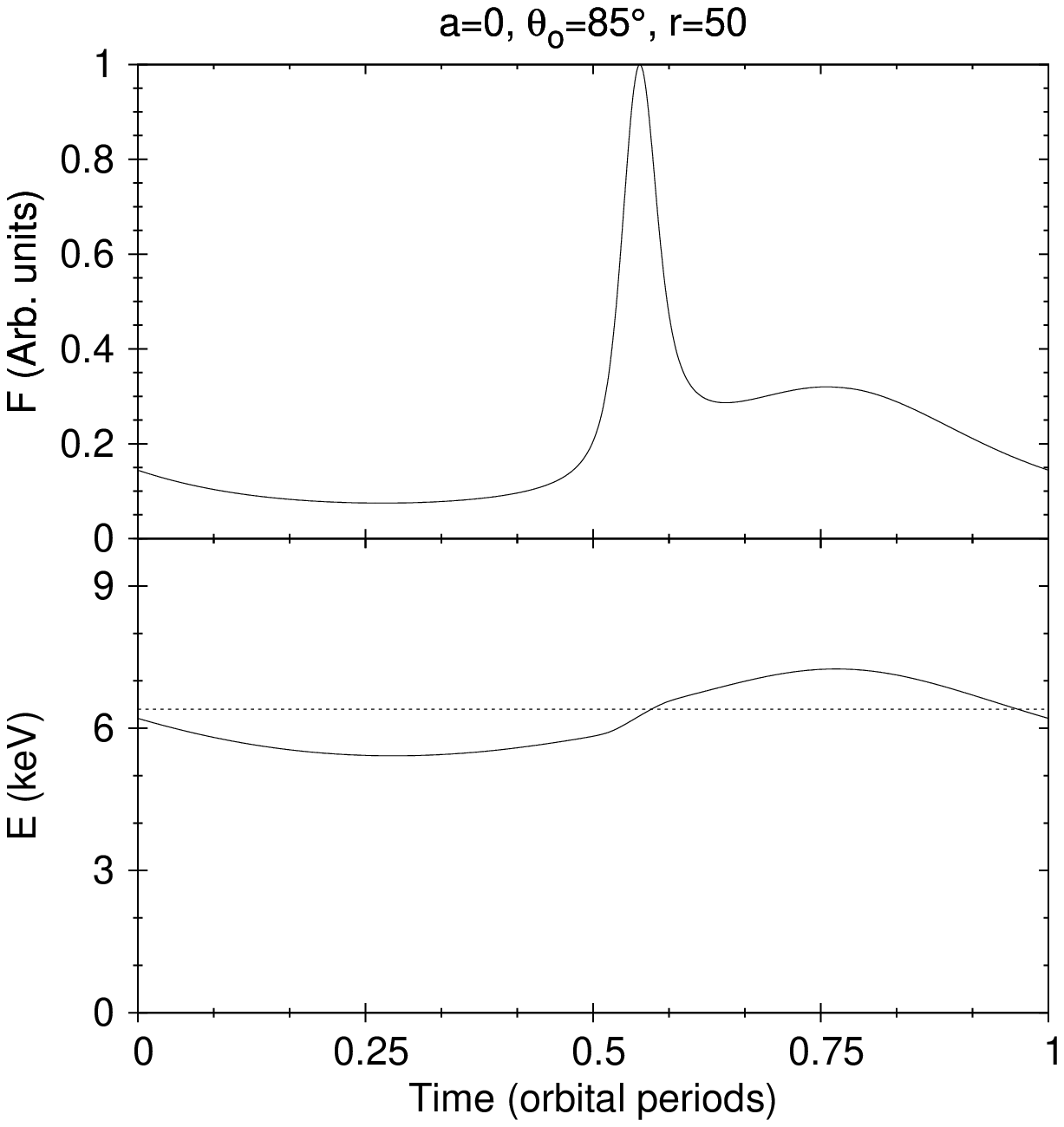}
\caption{The same as the previous figure, but with $a=0$ and
$r=10$ (top), $20$ (middle), and $50$ (bottom).}
\label{orbits_2}
\end{figure*}

\begin{figure*}
\includegraphics*[width=5.7cm,height=0.3\textheight]{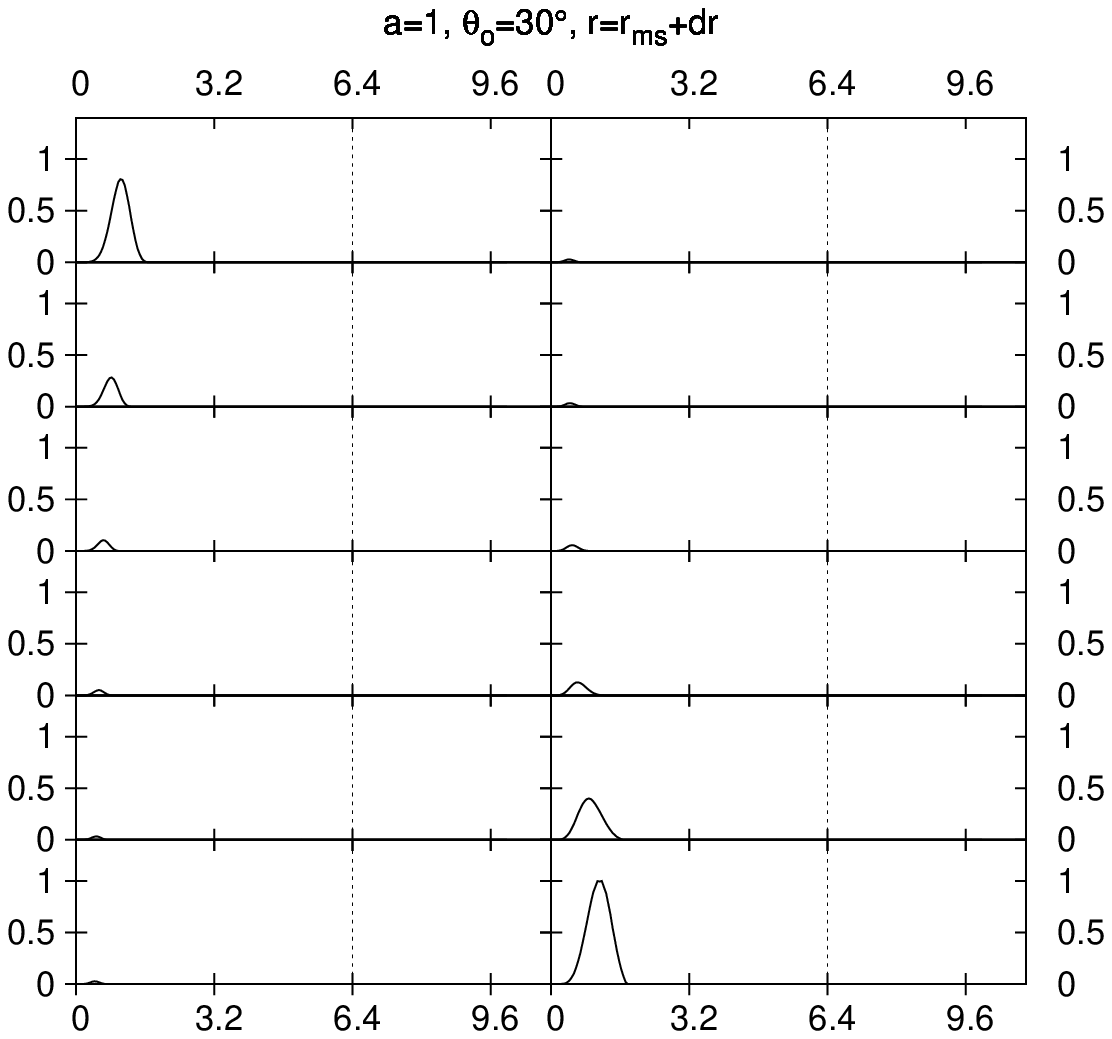}
\hfill
\includegraphics*[width=5.7cm,height=0.3\textheight]{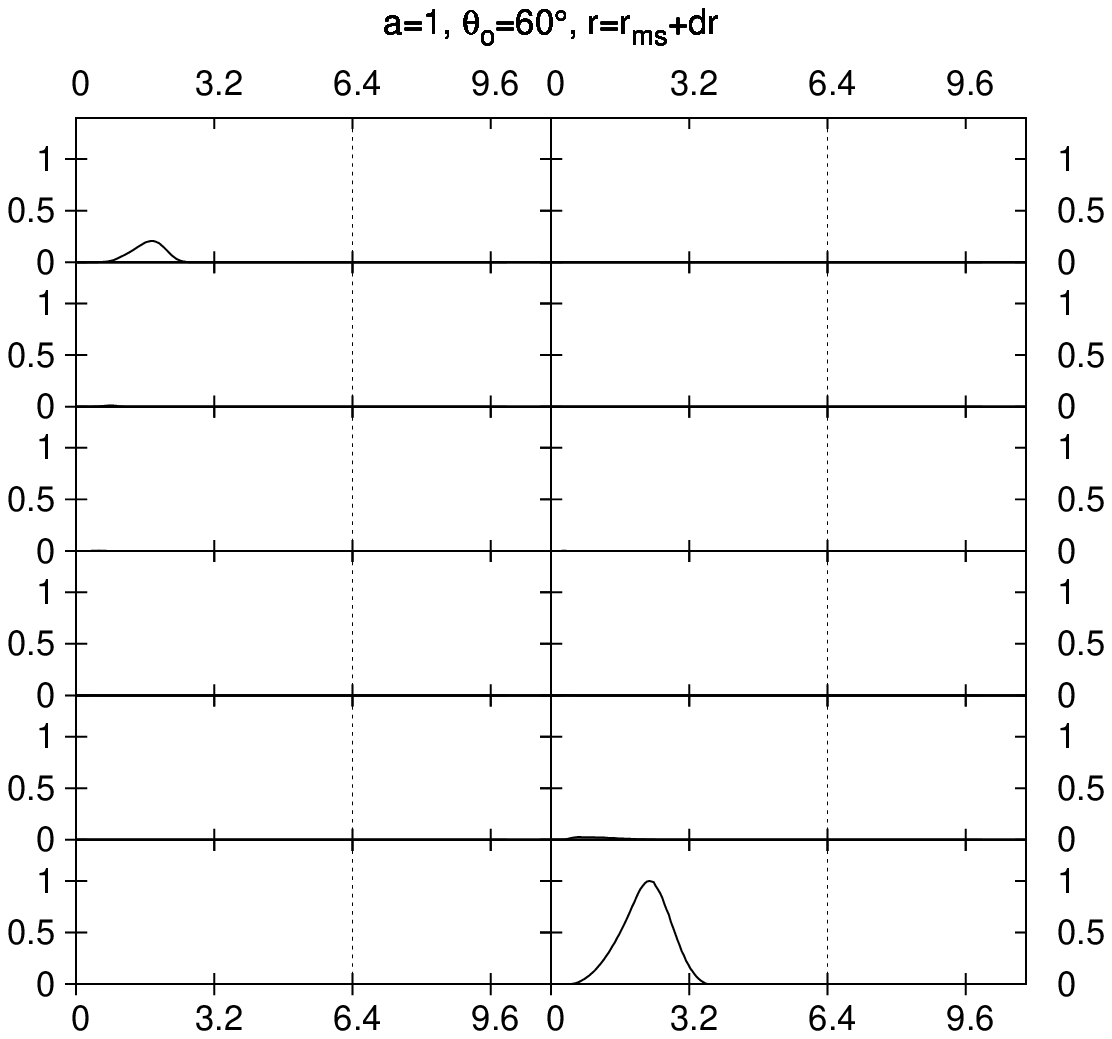}
\hfill
\includegraphics*[width=5.7cm,height=0.3\textheight]{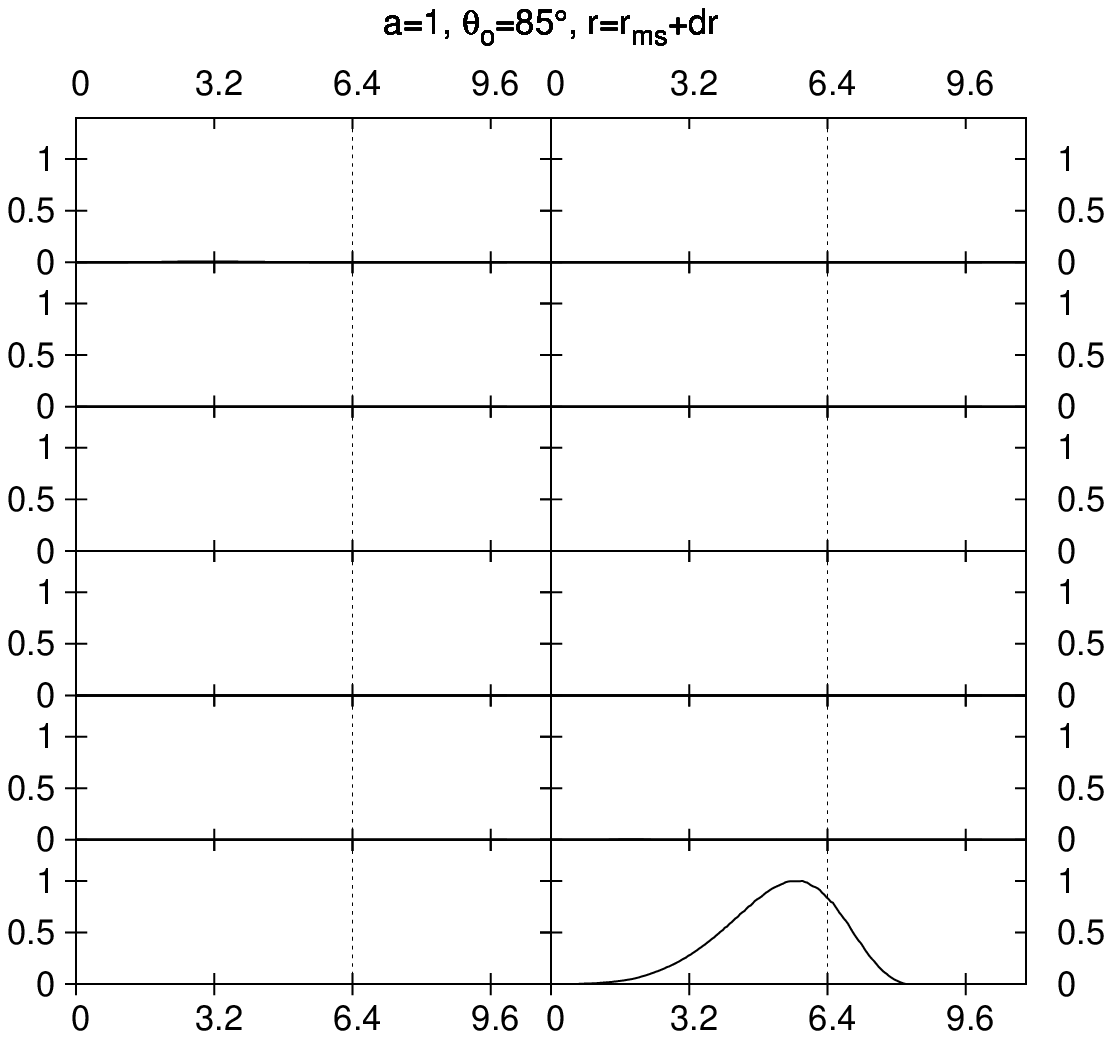}
\includegraphics*[width=5.7cm,height=0.3\textheight]{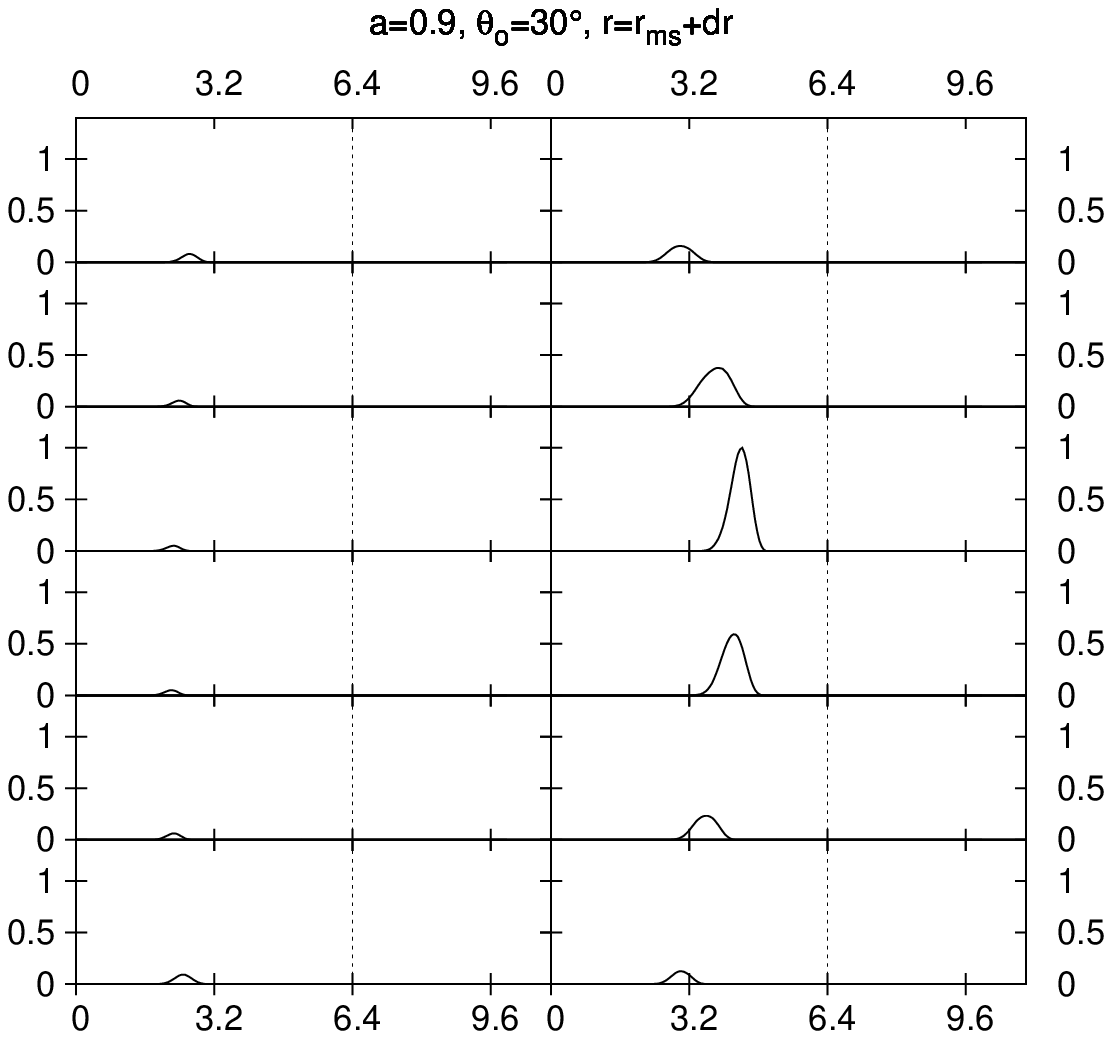}
\hfill
\includegraphics*[width=5.7cm,height=0.3\textheight]{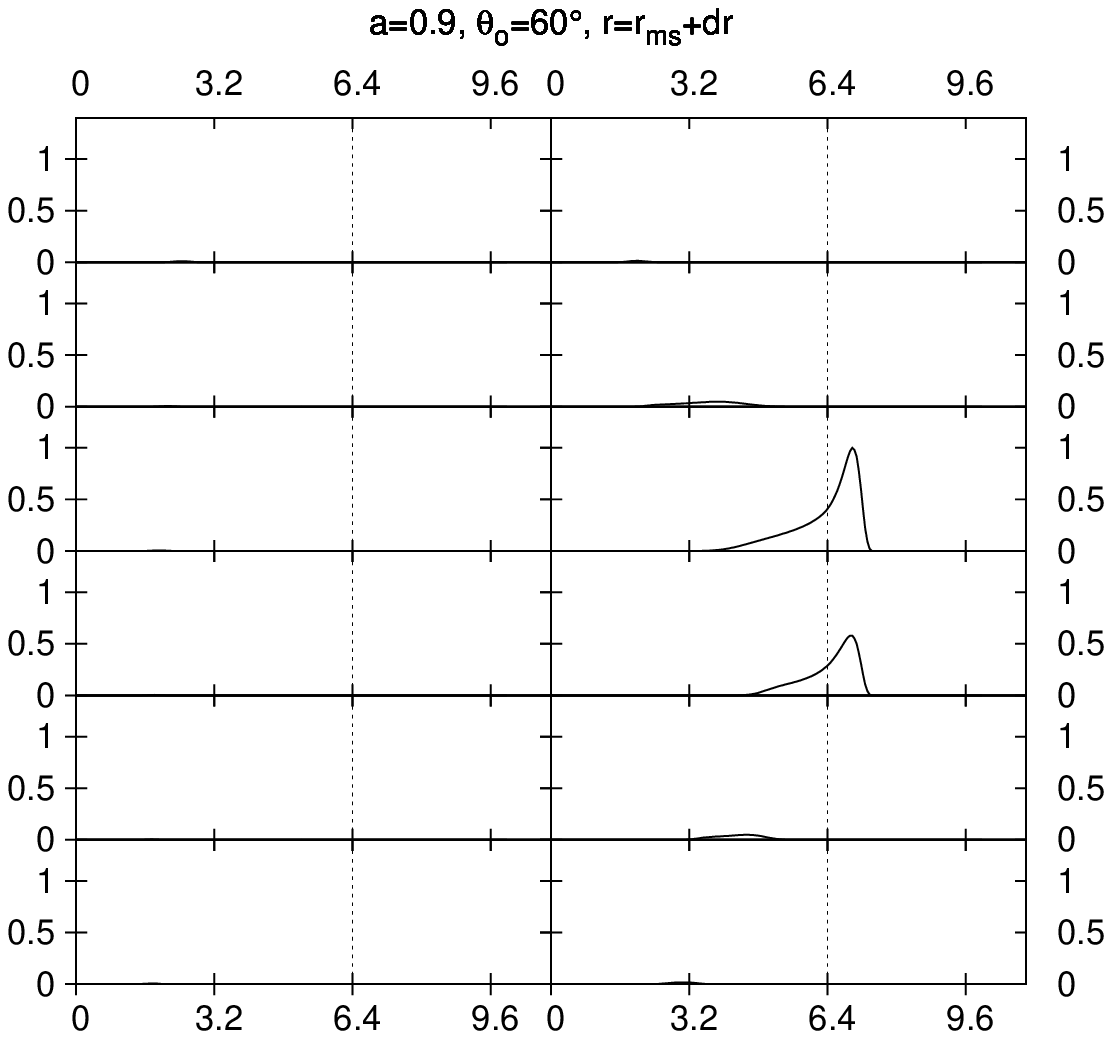}
\hfill
\includegraphics*[width=5.7cm,height=0.3\textheight]{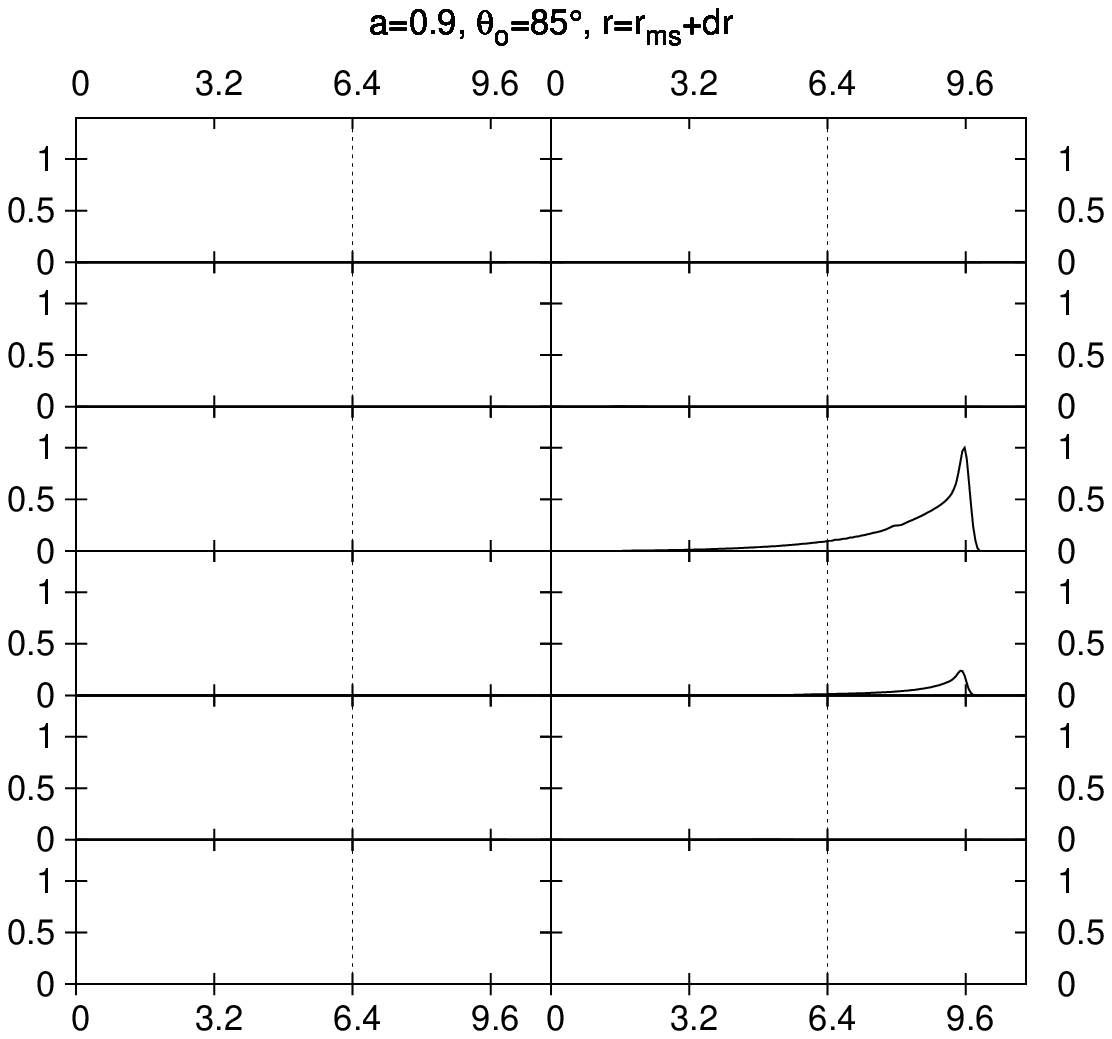}
\includegraphics*[width=5.7cm,height=0.3\textheight]{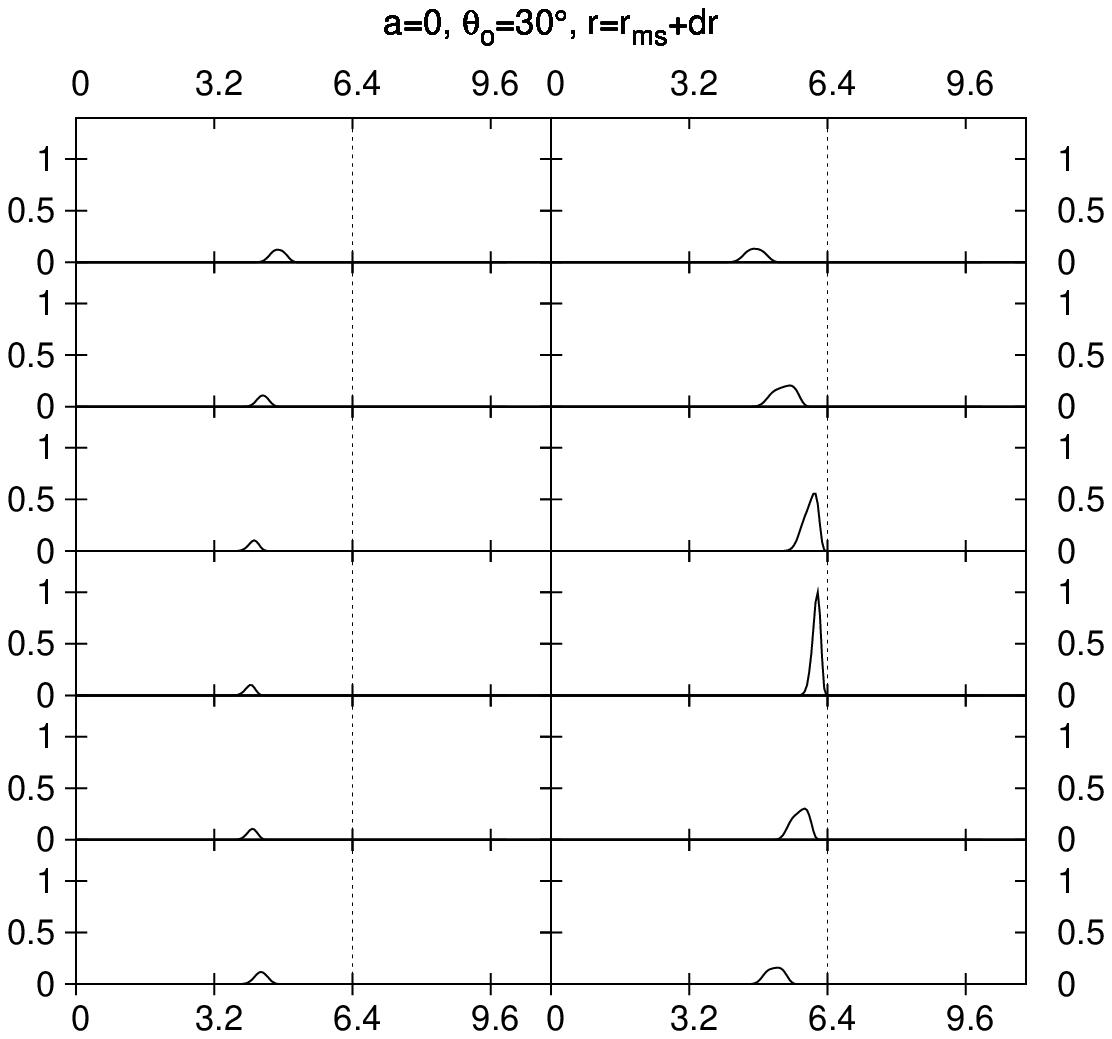}
\hfill
\includegraphics*[width=5.7cm,height=0.3\textheight]{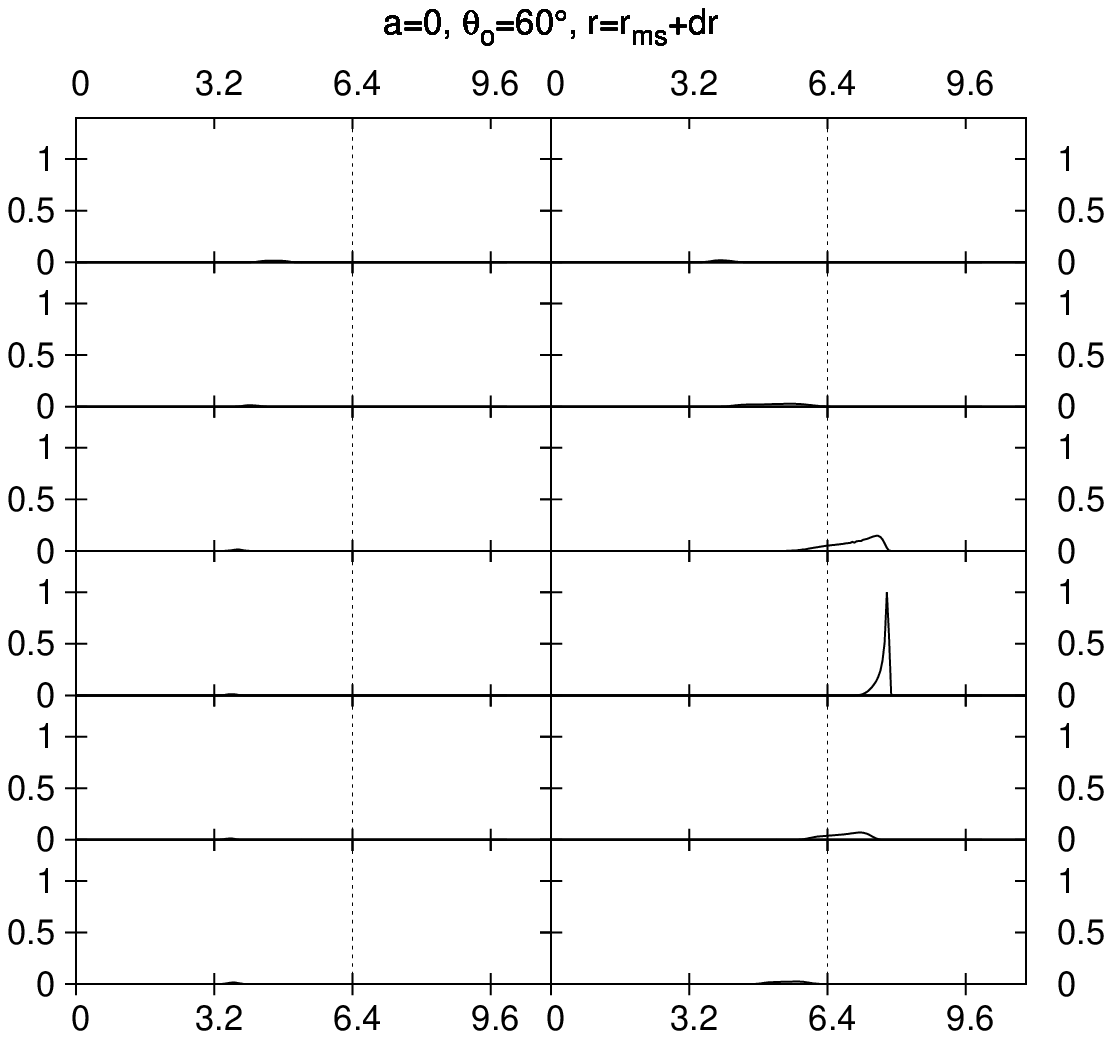}
\hfill
\includegraphics*[width=5.7cm,height=0.3\textheight]{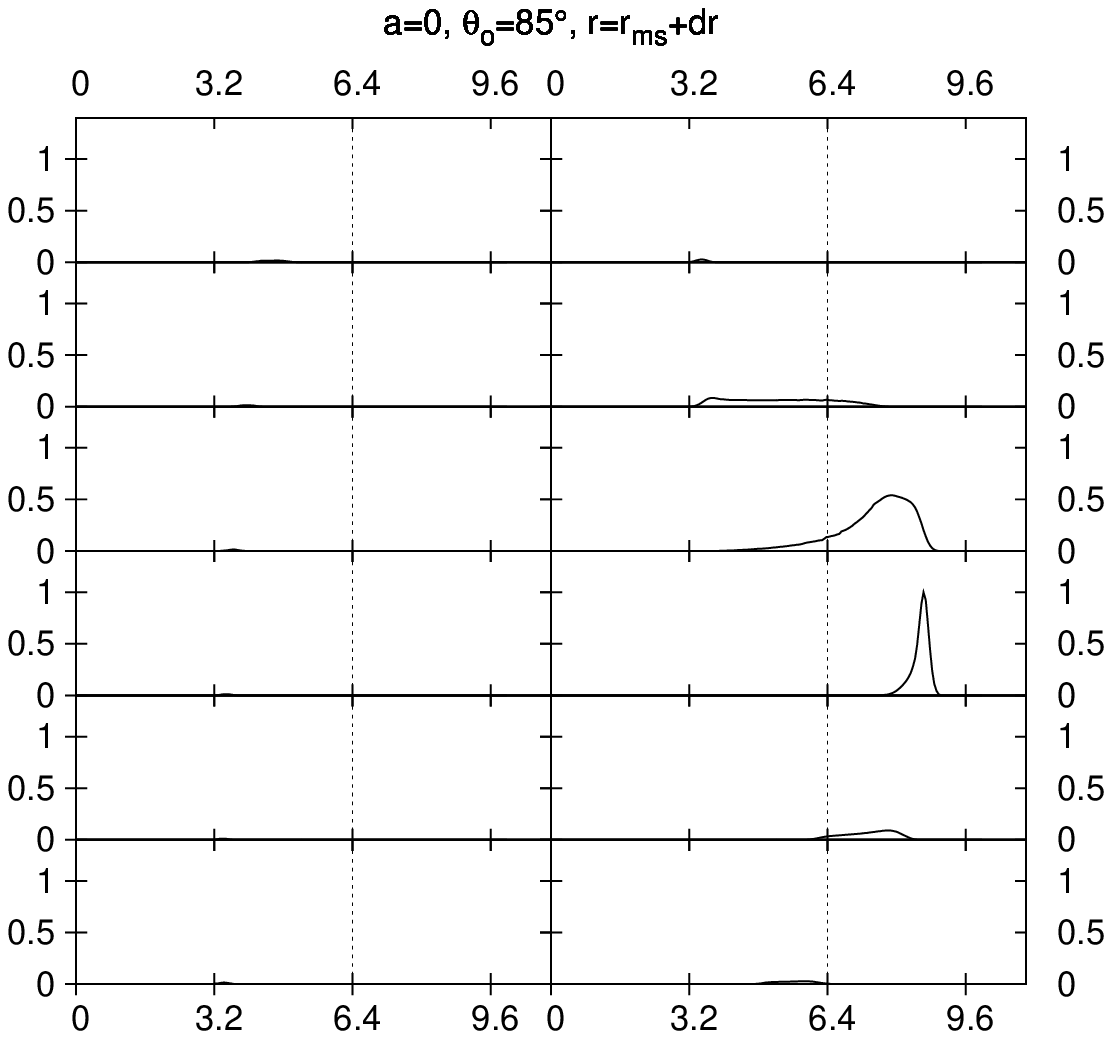}
\caption{Line profiles integrated over twelve consecutive
temporal intervals of equal duration. Each interval covers $1/12$ 
of the orbital period at corresponding radius. As explained in the text,
top-left frame of each panel corresponds to the spot being observed
at the moment of passing through lower conjunction. Energy is on abscissa 
(in keV). Observed photon flux is on ordinate (arbitrary units, scaled
to the maximum flux which is reached during the complete revolution
of the spot). Notice the occurrences of narrow and prominent peaks which 
appear for relatively brief fraction of the total period. Vertical
dotted lines indicate the line rest frame energy.}
\label{prof_phi_1}
\end{figure*}

\begin{figure*}
\includegraphics*[width=5.7cm,height=0.3\textheight]{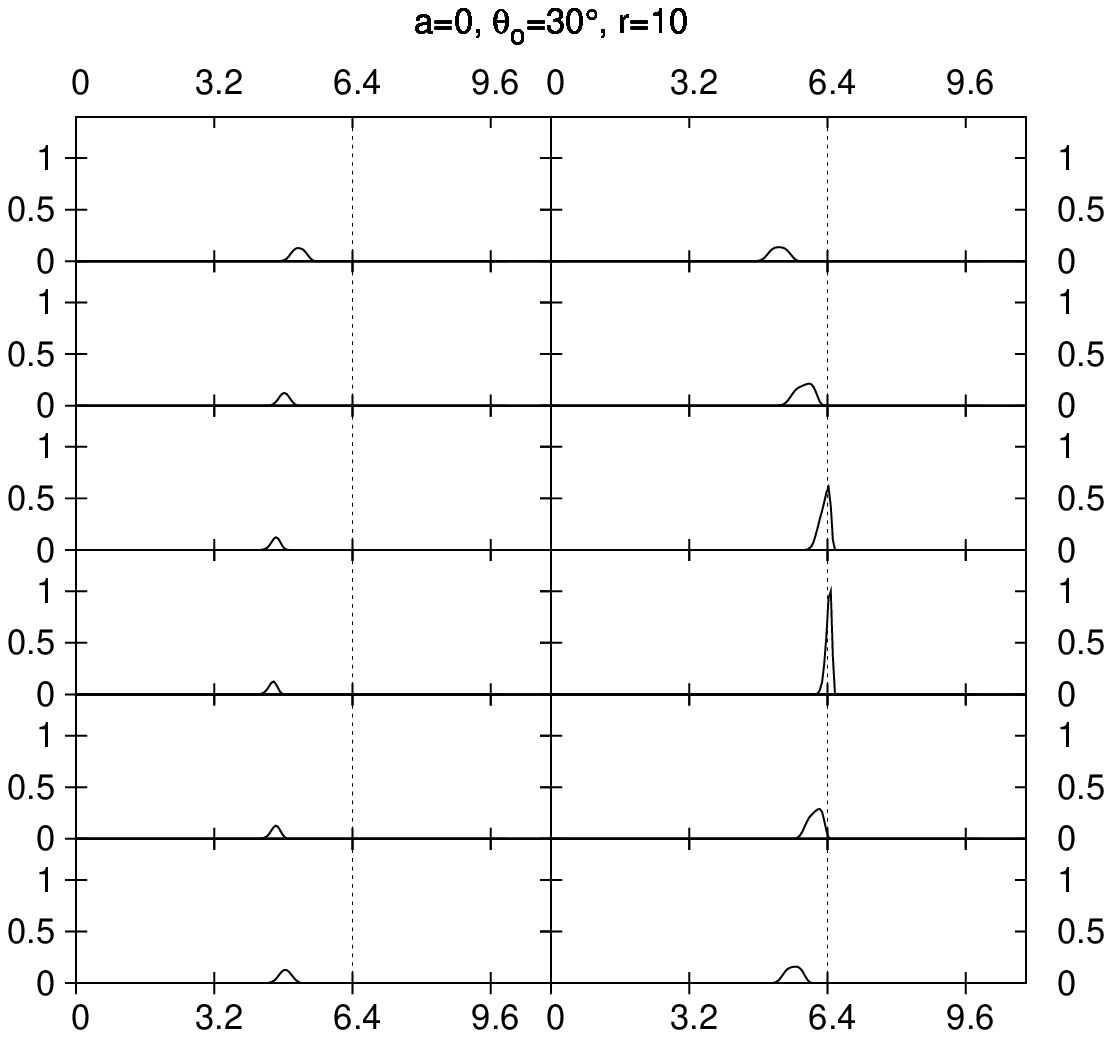}
\hfill
\includegraphics*[width=5.7cm,height=0.3\textheight]{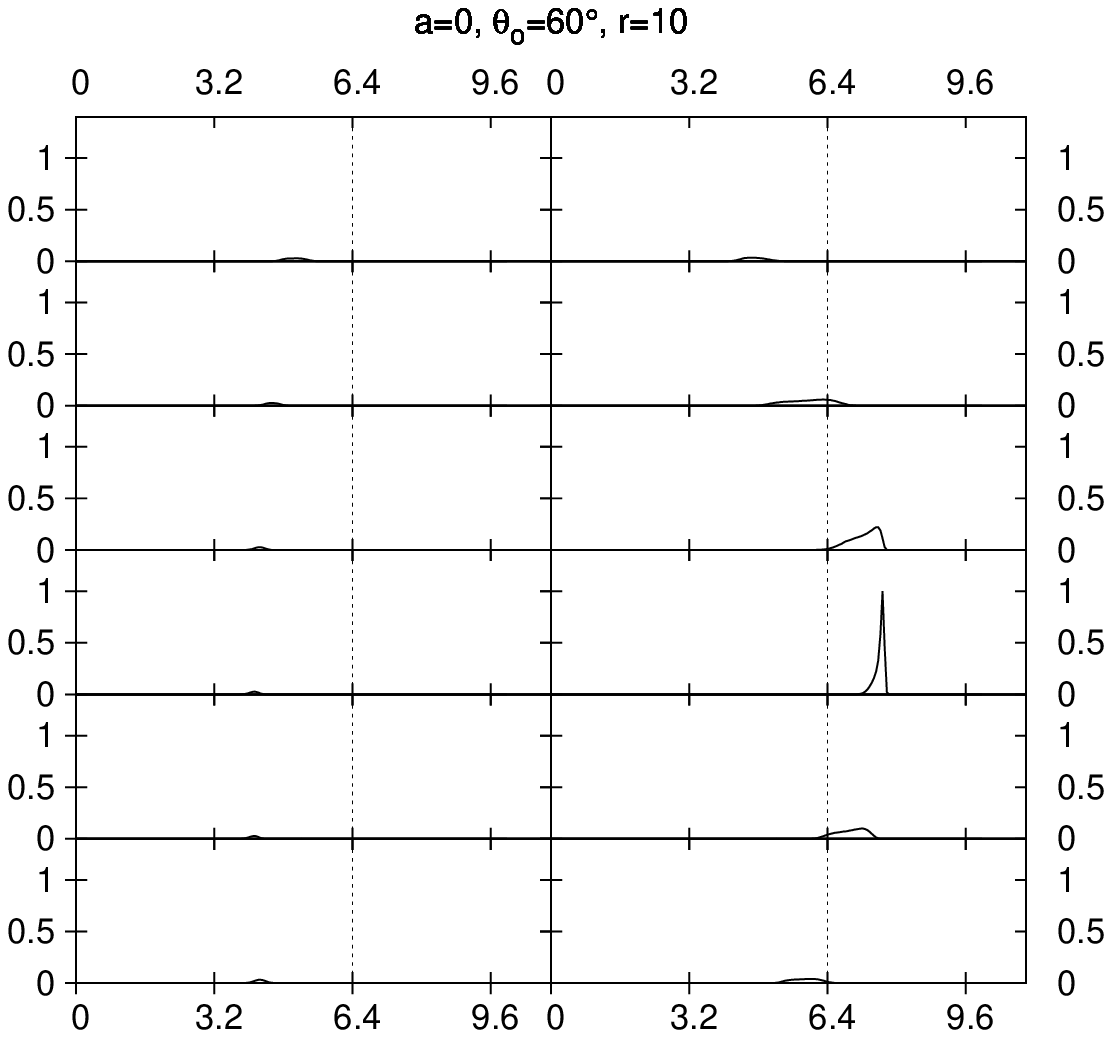}
\hfill
\includegraphics*[width=5.7cm,height=0.3\textheight]{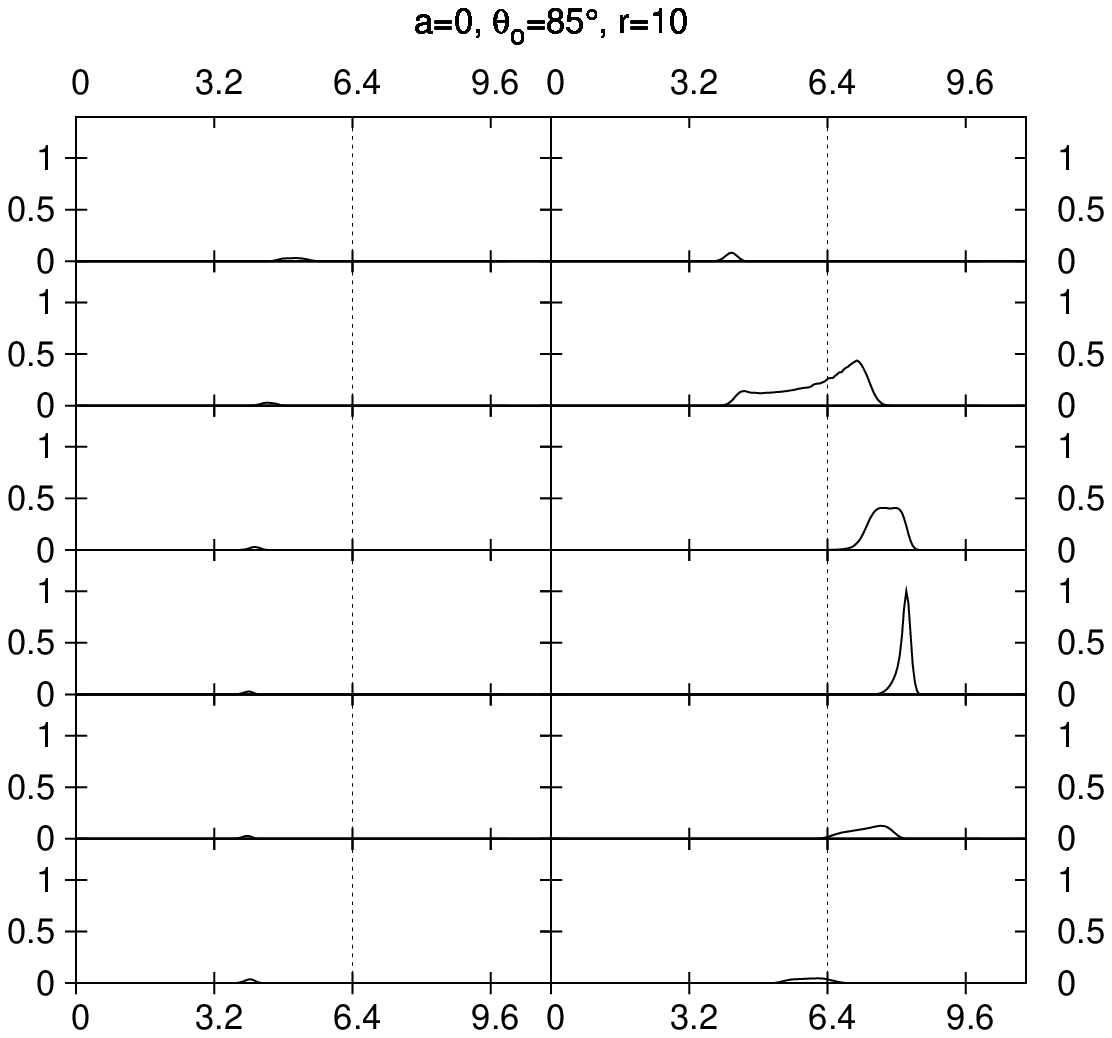}
\includegraphics*[width=5.7cm,height=0.3\textheight]{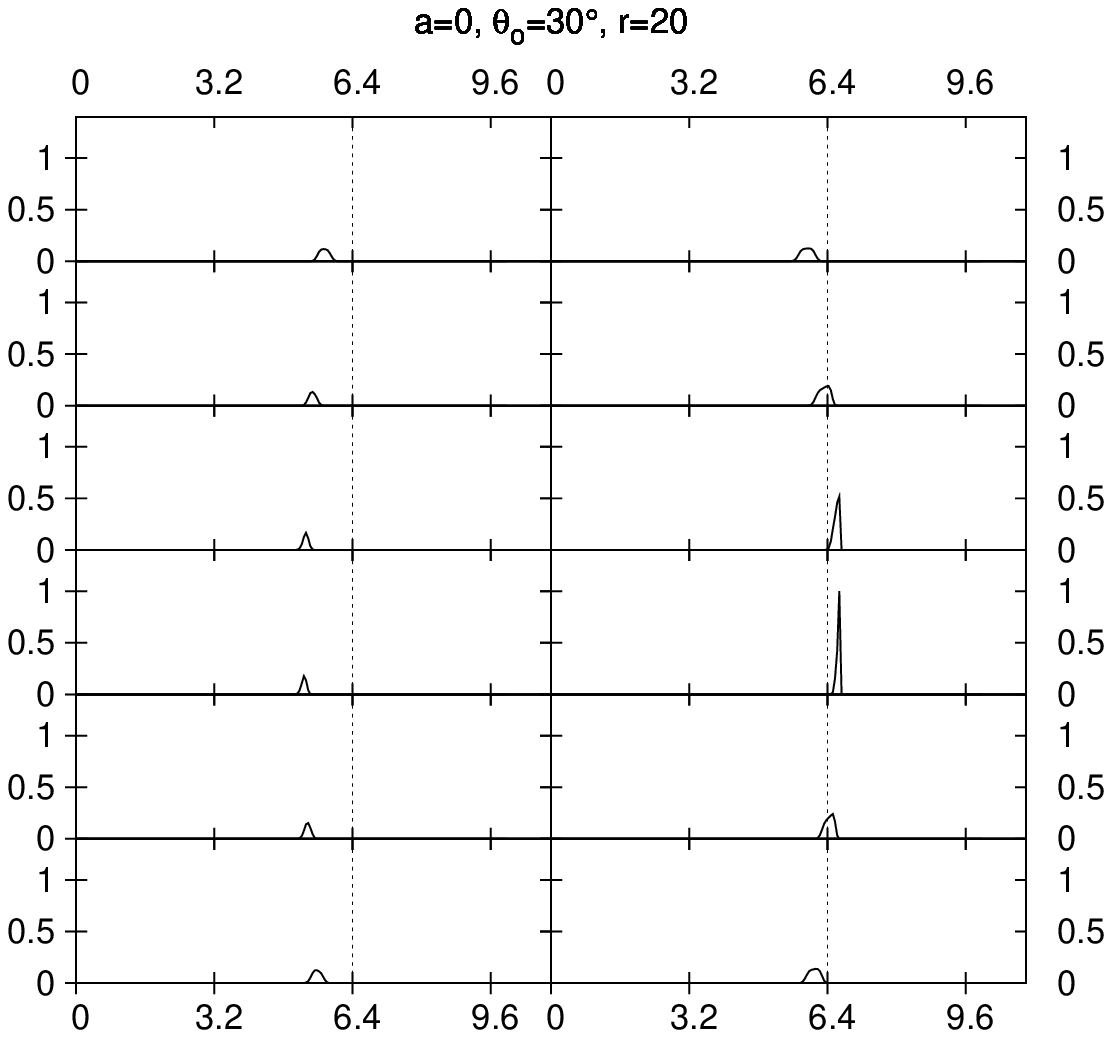}
\hfill
\includegraphics*[width=5.7cm,height=0.3\textheight]{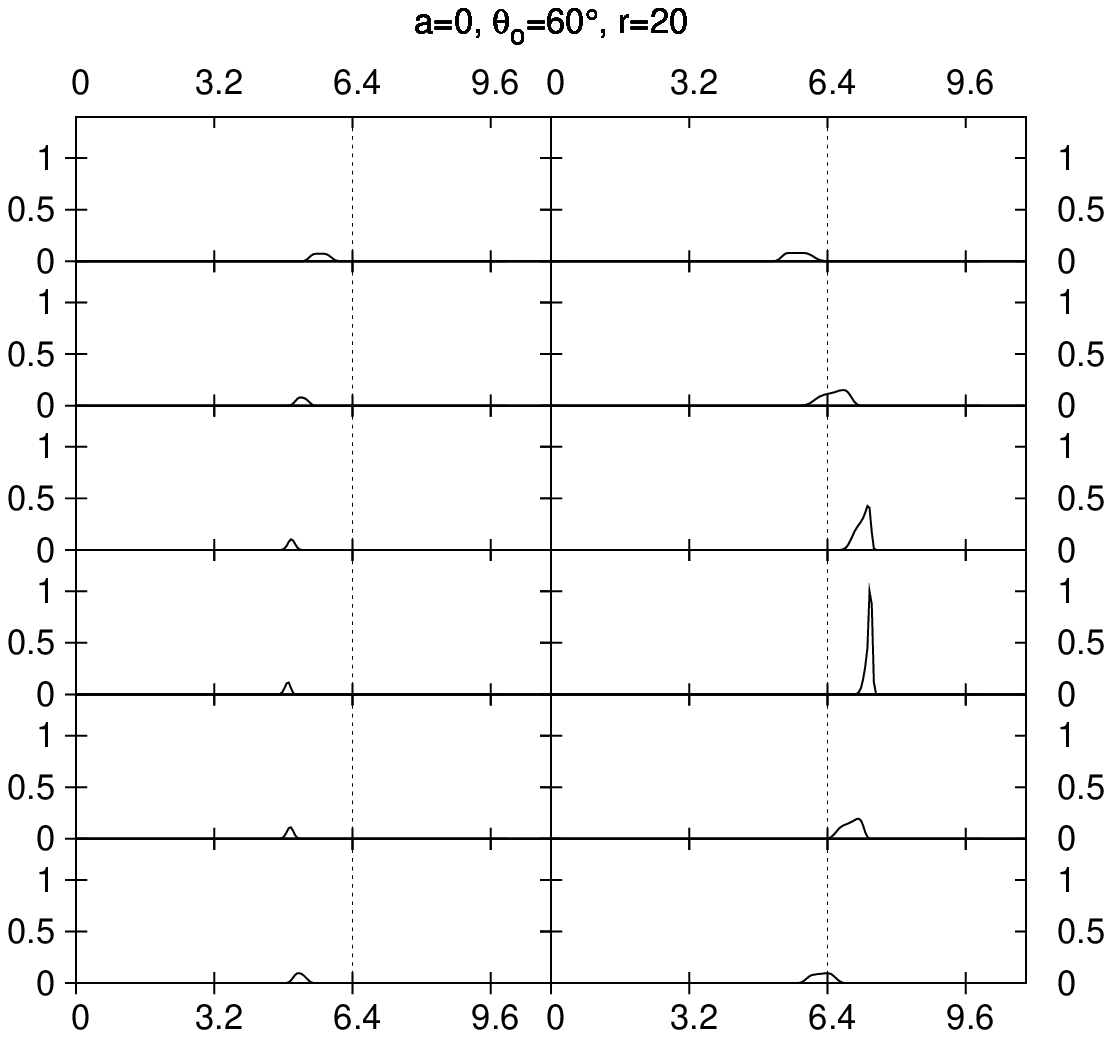}
\hfill
\includegraphics*[width=5.7cm,height=0.3\textheight]{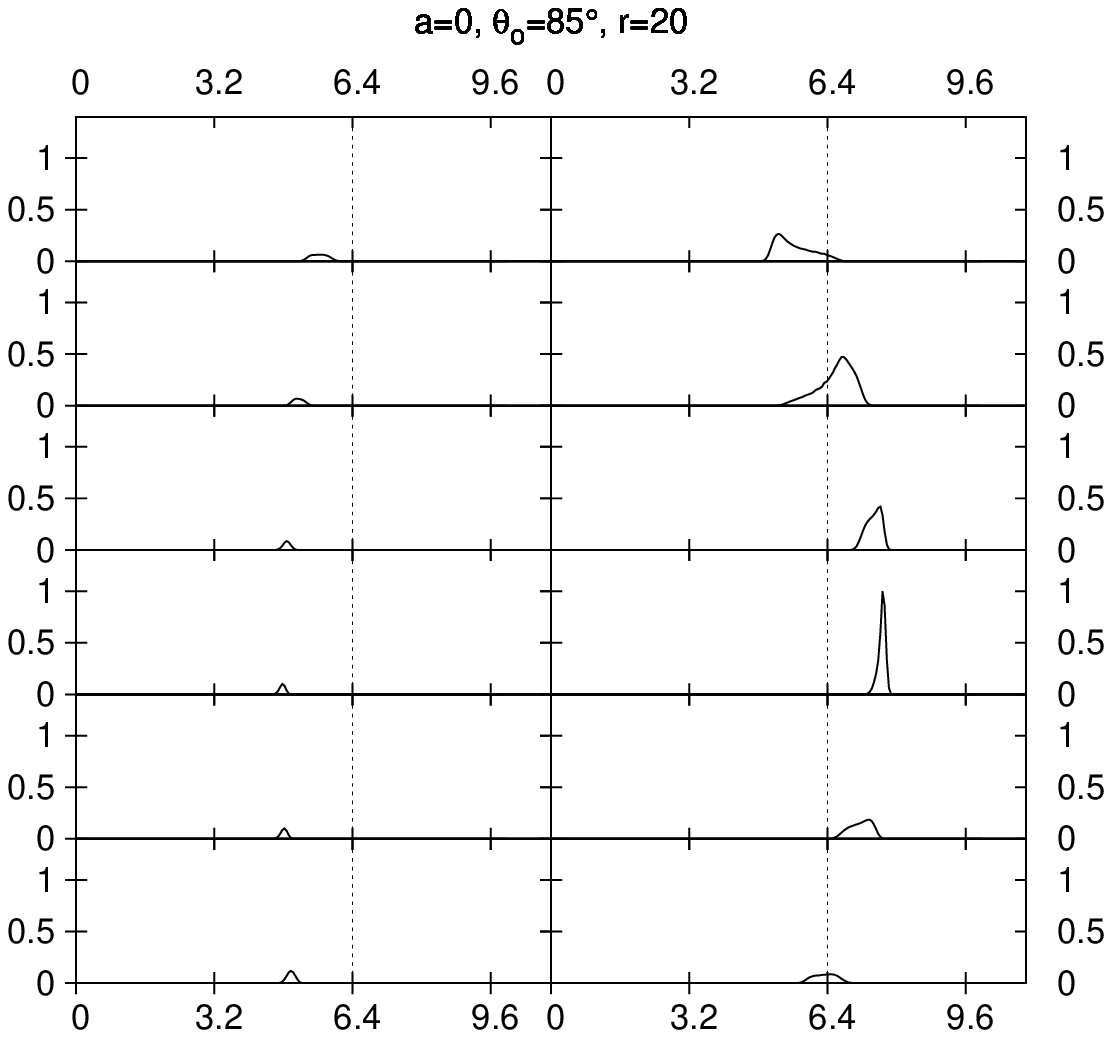}
\includegraphics*[width=5.7cm,height=0.3\textheight]{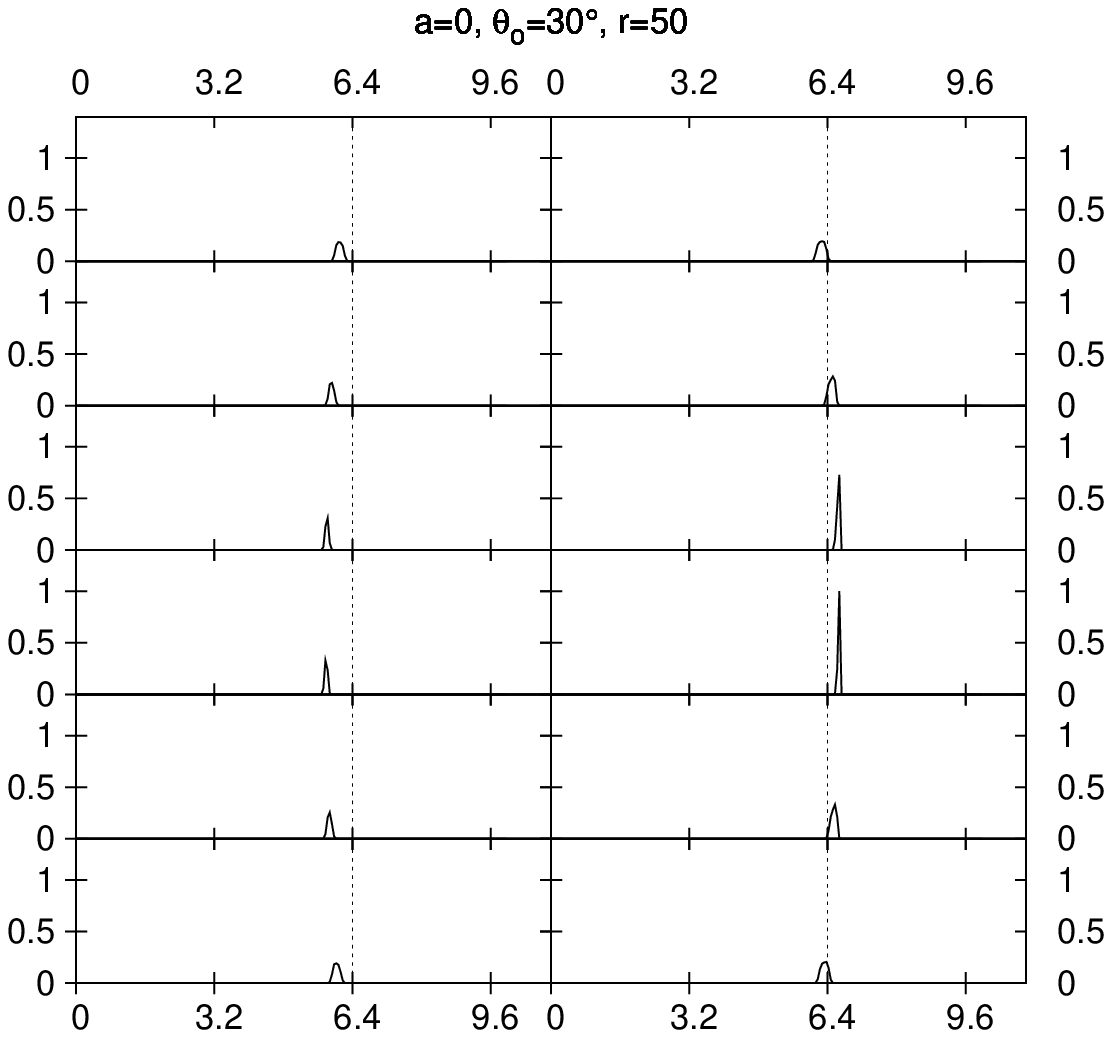}
\hfill
\includegraphics*[width=5.7cm,height=0.3\textheight]{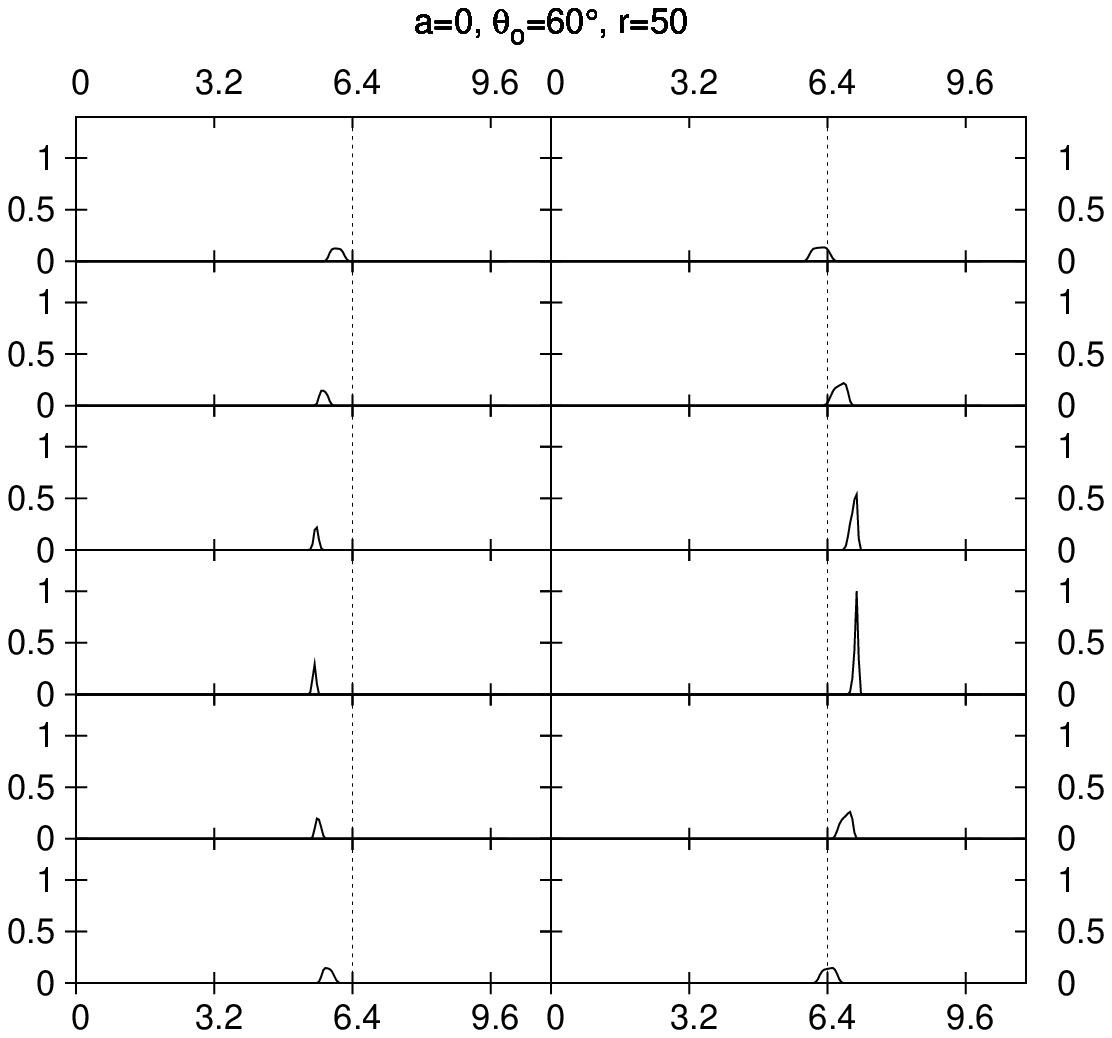}
\hfill
\includegraphics*[width=5.7cm,height=0.3\textheight]{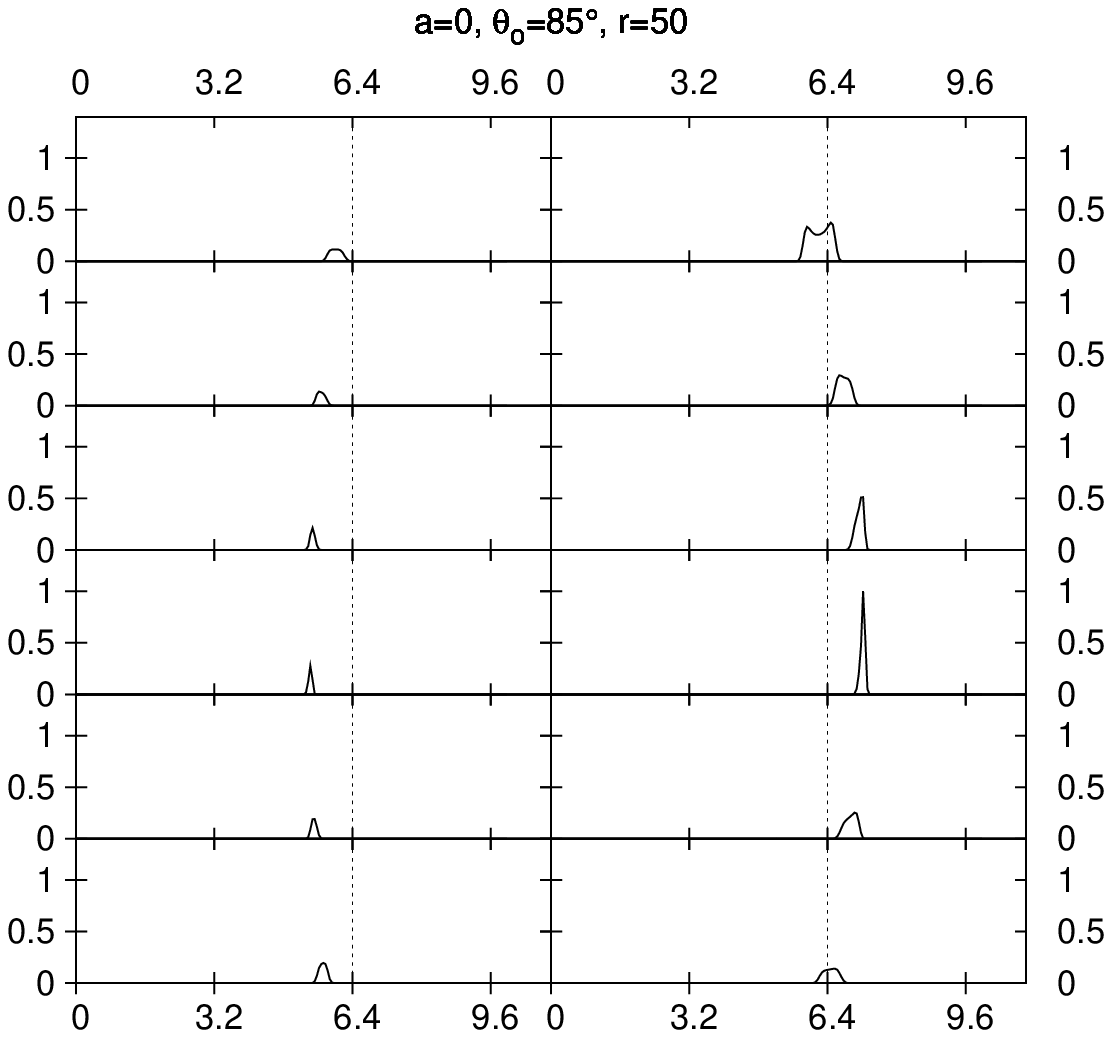}
\caption{The same as in previous figure, but with $r=10$ (top), $20$ 
(middle), and $50$ (bottom). The black hole was assumed non-rotating,
$a=0$ in this figure.}
\label{prof_phi_2}
\end{figure*}

\section{Line profiles from an orbiting, X-ray illuminated spot}
The basic properties of line emission from the innermost regions of an
accretion disc around a black hole are well-known (see e.g. Reynolds \&
Nowak 2003; Fabian et al. 2000 for recent reviews). Let us here briefly summarize several
formulae most relevant to our purposes.

If $r$ is the orbital radius and $a$ is the dimension-less black hole
angular momentum, the orbital period of matter co-rotating along a 
circular trajectory $r=\mbox{const}$ around the black hole 
is given by (Bardeen, Press \& Teukolsky 1972)
\begin{equation}
T_{\rm{}orb} \doteq 310~\left(r^\frac{3}{2}+a\right)
\frac{\mbh}{10^7M_{\odot}}\quad\mbox{[sec]},
\label{torb}
\end{equation}
as measured by a distant observer. We express lengths
in units of the gravitational radius 
$r_{\rm{}g}{\equiv}G{\mbh}/c^2{\doteq}1.48\times10^{12}M_7$~cm, where
$M_7$ is the mass of the black hole in units of $10^7$ solar masses.
Angular momentum $a$ (per unit mass) is in geometrized units
($0\leq{a}\leq1$). See e.g. Misner, Thorne
\& Wheeler (1973) for useful conversion formulae between geometrized 
and physical units.

The innermost stable orbit, 
$r_{\rm ms}$, occurs for an equatorial disc at radius
\begin{equation}
r_{\rm ms} = 3+Z_2-\big[\left(3-Z_1)(3+Z_1+2Z_2\right)\big]^{1 \over 2},
\label{velocity1}
\end{equation}
where 
$Z_1 = 1+(1-a^2)^{1 \over 3}[(1+a)^{1 \over 3}+(1-a)^{1 \over 3}]$
and $Z_2 = (3a^2+Z_1^2)^{1 \over 2}$;
$r_{\rm ms}$ spans the range of radii from $r=1$ ($a=1$, i.e. the case
of a maximally rotating black hole) to $6$ 
($a=0$, a static black hole). Rotation of a black hole
is believed to be limited by an equilibrium value 
$a\dot{=}0.998$ because of the capture of photons from the disc (Thorne 1974). 
This would imply $r_{\rm{}ms}\dot{=}1.23$. Different models
of accretion can result in somewhat different limiting values
of $a$ and the corresponding $r_{\rm{}ms}(a)$. Notice that in the static case, 
the radial dependence $T_{\rm{}orb}(r)_{{\mid}a=0}$ is identical to that 
in purely Newtonian gravity. 

In order to compute a synthetic profile of an observed spectral line
one has to link the points of emission in the disc with corresponding
pixels in the detector plane at spatial infinity. This can be achieved by
solving the ray-tracing problem in curved space-time of the black hole.
Appropriate methods were discussed by several authors; see Reynolds
\& Nowak (2003) for a recent review and for further references.
This way one finds the redshift factor, which determines the
energy shift of photons, the lensing effect (i.e. the change of solid angle
due to strong gravity), and the effect of aberration (which influences
the emission direction of photons from the disc; this must be taken 
into account if the intrinsic emissivity is non-isotropic). We consider these
effects in our computations, assuming a rotating (Kerr)
black hole spacetime (Misner et al. 1973). We also consider time of arrival
of photons originating at different regions of the disc plane.
Variable travel time results in mutual time delay between different 
photons, which can be ignored when analyzing time-averaged data
but it may be important for time-resolved data.

Assuming purely azimuthal Keplerian motion of a spot, one obtains
for its orbital velocity (with respect to a locally non-rotating
observer at corresponding radius $r$):
\begin{equation}
 v^{(\phi)} = \frac{r^2-2a\sqrt{r}+{a}^2}{\sqrt{\Delta}\left(r^{3/2}
 +{a}\right)}.
\end{equation}
In order to derive time and frequency as measured by a distant observer,
one needs to take into account the Lorentz factor associated with
this orbital motion,
\begin{equation}
 \Gamma = \frac{\left(r^{3/2}+{a}\right)\sqrt{\Delta}}{r^{1/4}\;
  \sqrt{r^{3/2}-3r^{1/2}+2{a}}\;\sqrt{r^3+{a}^2r+2{a}^2}}\,.
\end{equation}
Corresponding angular velocity of orbital motion is
$\Omega=(r^{3/2}+{a})^{-1}$, which also
determines the orbital period in eq.~(\ref{torb}).
The redshift factor $g$ and the emission angle $\vartheta$ (with respect to
the normal direction to the disc) are then given by
\begin{equation}
 g  =  \frac{{\cal{C}}}{{\cal{B}}-r^{-3/2}\xi}, 
 \quad
 \vartheta  =  \arccos\frac{g\sqrt{\eta}}{r},
\end{equation}
where ${\cal{B}}=1+{a}r^{-3/2}$,
${\cal{C}}=1-3r^{-1}+2{a}r^{-3/2}$; $\xi$ and $\eta$ are constants 
of motion connected with the photon ray in an axially symmetric
and stationary spacetime.

For practical purposes formula (\ref{torb}) with $a=0$ is accurate 
enough also in the case of a spinning black hole,
provided that $r$ is not
very small. For instance, even for $r=6$ (the last stable orbit in
Schwarzschild metric), $T_{\rm orb}(r_{\rm ms})$ calculated for a 
static and for a maximally rotating ($a=1$) black hole differ by 
about $6.8$\%. The relative difference decreases, roughly linearly, down to
$1.1$\% at $r=20$. 
This implies that eq.~(\ref{torb}) can be used
in most cases to estimate the black hole mass even if the angular momentum
is not known (deviations are relevant only for $r<6$, when the radius itself can be 
used to constrain the allowed range of $a$). 

Various
pseudo-Newtonian formulae have been devised for accreting black holes
to model their observational properties, which are connected with
the orbital motion of surrounding matter (e.g. Abramowicz et al. 1996; 
Artemova, Bj\"{o}rnsson \& Novikov 1996; Semer\'ak \& Karas 1999). 
Although this approach
is often used and found to be practical, we do not employ it here because 
error estimates are not possible within the pseudo-Newtonian scheme.

Due to Doppler and gravitational energy shift the line shape changes along 
the orbit. Centroid energy is redshifted with respect to the rest
energy of the line emission for most of the orbit.
Furthermore, light aberration and bending cause the flux to
be strongly phase--dependent. These effects are shown in
Figures \ref{orbits_1}--\ref{orbits_2}. In these plots, 
the arrival time of photons is defined in orbital periods, 
i.e.\ scaled with $T_{\rm{}orb}(r;a)$. The
orbital phase of the spot is of course linked with the
azimuthal angle in the disc, but the relation is made complex
by time delays which cannot be neglected, given the large velocities of 
the orbiting matter and frame-dragging effects near the black hole. 
Here, zero time corresponds to the moment
when the center of the spot was at the nearest point on its orbit with
respect to the observer (a lower conjunction).
The plots in Fig.~\ref{orbits_1} refer to the case of a spot
circulating at the innermost stable
orbit $r_{\rm{}ms}(a)$ for $a=0$, $0.9$ and $1$.
The effect of black hole rotation becomes prominent for almost
extreme values of $a$; one can check, for example, that the
difference between cases $a=0$ and $a=0.5$ is very small.
 
Worth remarking is a large difference in the orbital phase of 
maximum emission between the extreme case, $a\rightarrow1$, in 
contrast to the non-rotating case, $a\rightarrow0$. The reason is 
that for large $a$ the time delay and the effect of frame-dragging
on photons emitted behind the black hole are very substantial. It
is also interesting to note that, for very high inclination angles, most
of the flux comes from the far side of the disc, due to very strong
light bending, as pointed out by Matt et al.\ (1992, 1993) and 
examined further by many authors who performed detailed ray tracing,
necessary to determine the expected variations of the line flux and 
shape. A relatively simple fitting formula has been also derived 
(Karas 1996) and can be useful for practical computations.

Three more orbits (centered at $r=10$, $20$ and $50$) are shown for 
$a=0$ (Figure~\ref{orbits_2}). As said above, at
these radii differences between spinning and static black holes are
small. Indeed, it can be verified that the dependence on $a$ is only 
marginal if $r\ga20$, and so it can be largely neglected for 
present-day measurements. 

In Figures~\ref{prof_phi_1}--\ref{prof_phi_2} we show
the actual form of the line profiles for the same sets of parameters as 
those explored in Figs.~\ref{orbits_1}--\ref{orbits_2}. The entire
revolution was split into
twelve different phase intervals. The intrinsic flux $I$ is
assumed to decrease exponentially with the distance ${\rm{}d}r$ 
from the centre of the spot
(i.e. ${\log}I\propto-[\kappa\,{\rm{}d}r/r]^2$, where $r$ is the 
location of the spot centre and $\kappa\sim10$ is a constant). 
The illumination is supposed to cease at distance ${\rm{}d}r$ 
away from the spot centre, which also defines the illuminated area in the 
disc. Let us remark that we concentrate on a spectral line
which is intrinsically narrow and unresolved in the rest frame of
the emitting medium. Such a line can be produced by a spot which 
originates due to sharply
localized illumination by flares, as proposed and discussed
by various authors (e.g. Haardt, Maraschi \& Ghisellini 1994;
Poutanen \& Fabian 1999; Merloni \& Fabian 2001). 

Very recently, Czerny et al. (2004) have examined
the induced {\sf{}rms} variability in the flare/spot 
model with relativistic effects. In this scheme, the actual size
of the spot is linked with the X-ray flux, which is produced in the flare,
and with the vertical height at which the flare occurs above the disc plane.
These quantities are obviously model-dependent. Czerny et al. (2004) 
computations provided different cases with the spot size ranging from
a fraction of $r_{\rm{}g}$ to several units of $r_{\rm{}g}$. Our 
computations can also simulate the observed features with the spot size
as a free parameter, but with present data 
we cannot constrain this parameter with sufficient accuracy. 
Obviously, the idea of narrow spectral features favours small
size of the spot, as large spots would produce broader features
and they would be more prone to rapid destruction. 
Hence, we fix the spot size somewhat arbitrarily at a lower
boundary, ${\rm{}d}r=0.2r$.

\begin{figure*}
\includegraphics*[width=5.3cm]{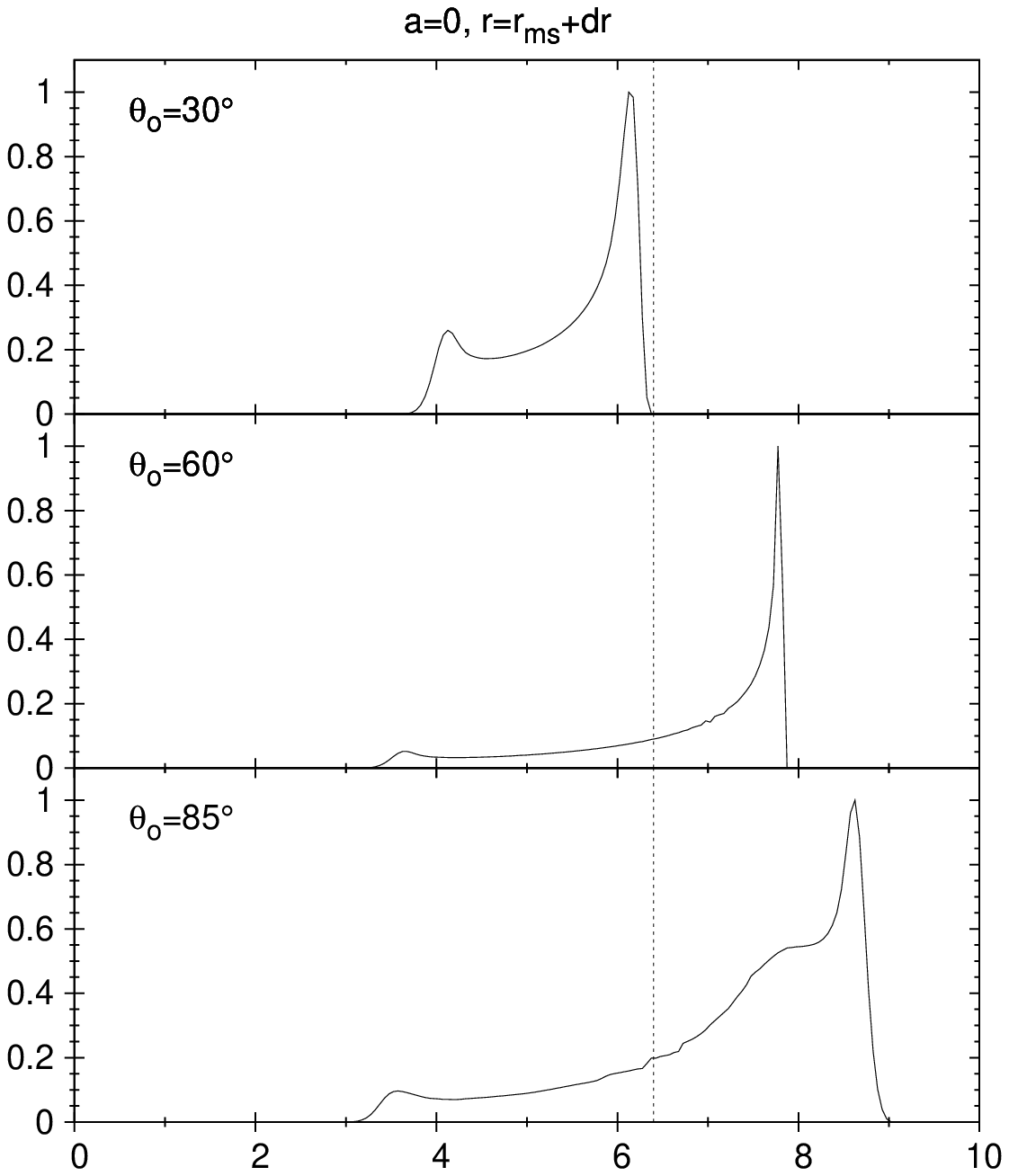}
\hfill
\includegraphics*[width=5.3cm]{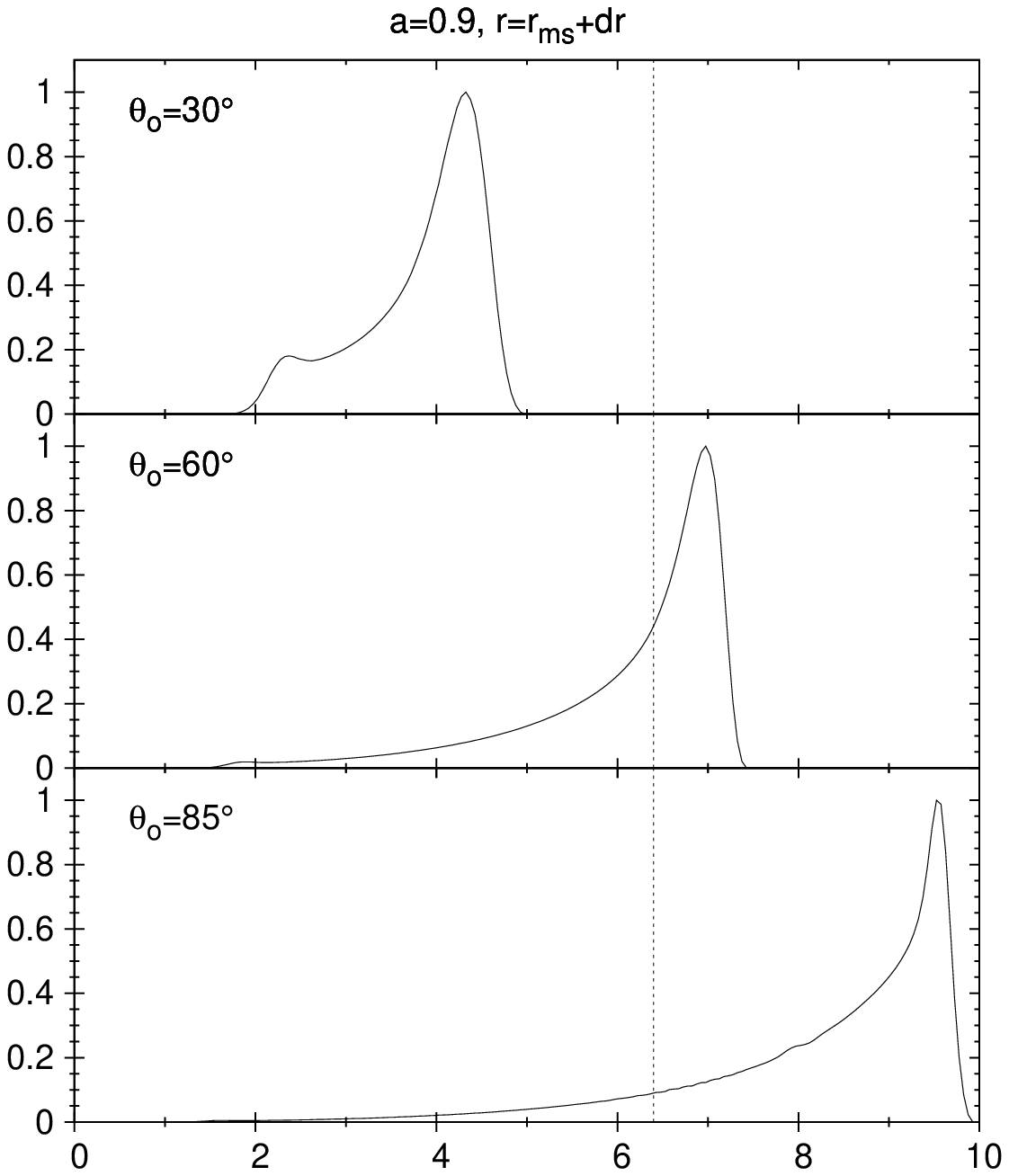}
\hfill
\includegraphics*[width=5.3cm]{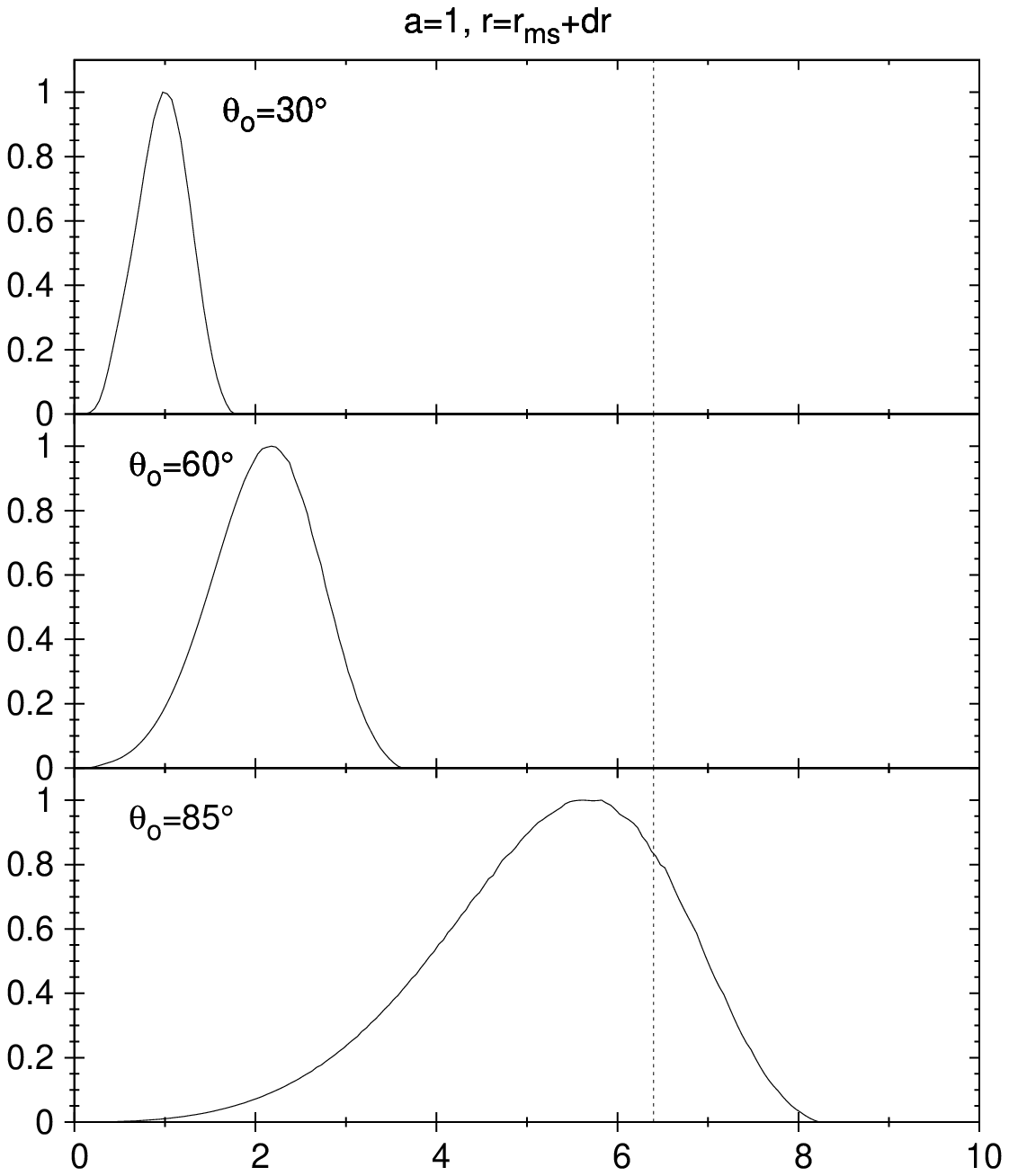}
\vspace*{3mm}\\
\includegraphics*[width=5.3cm]{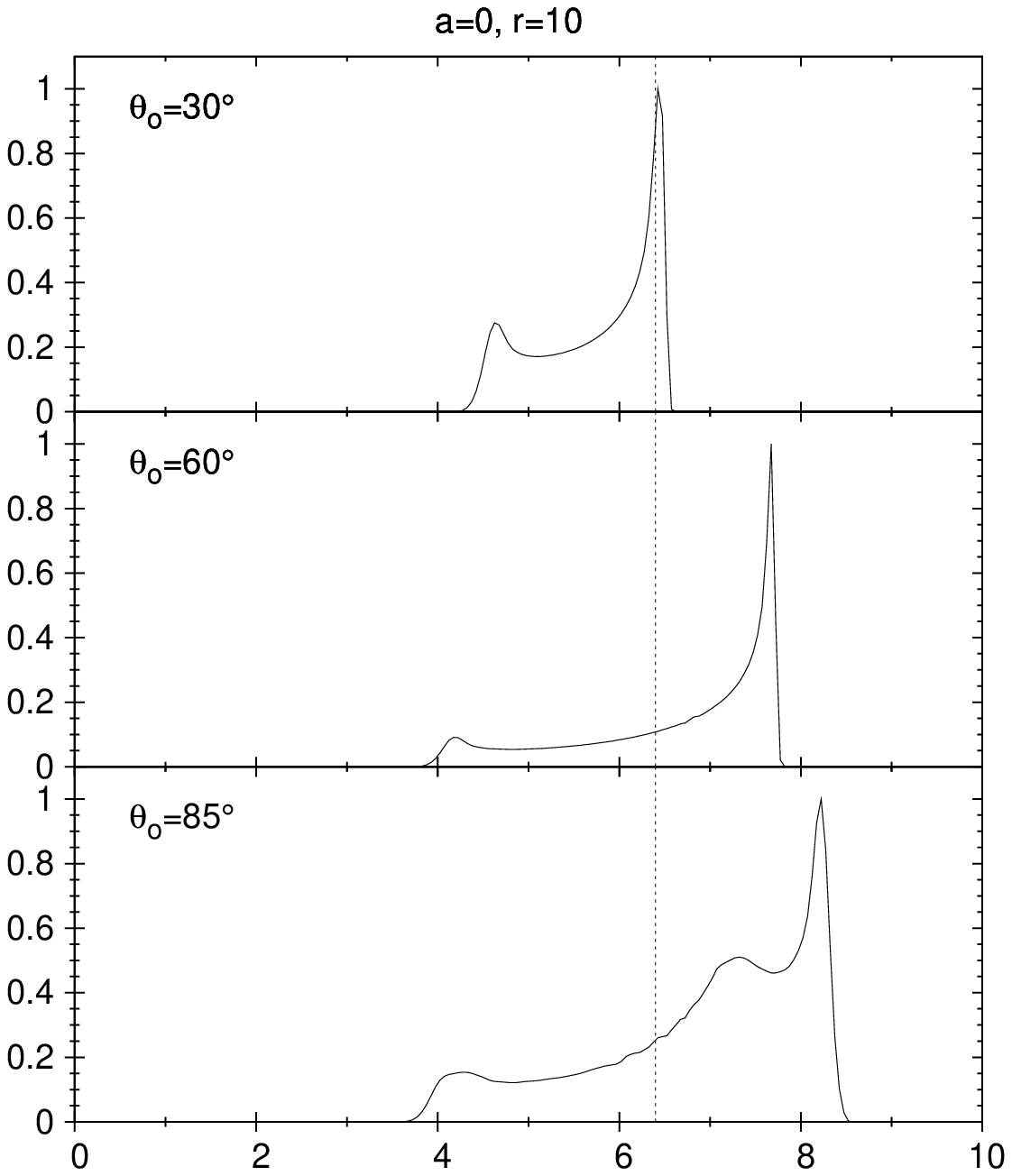}
\hfill
\includegraphics*[width=5.3cm]{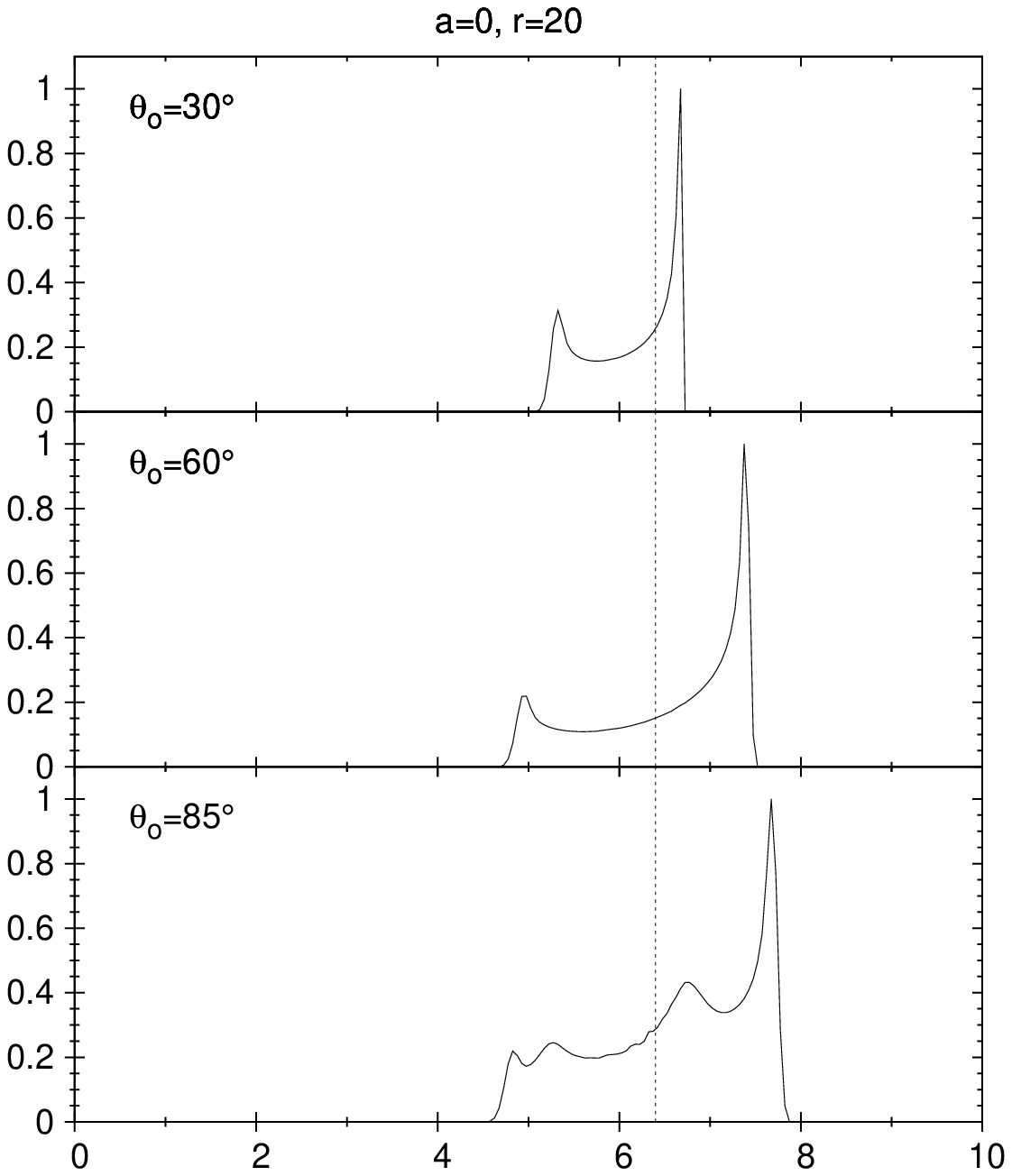}
\hfill
\includegraphics*[width=5.3cm]{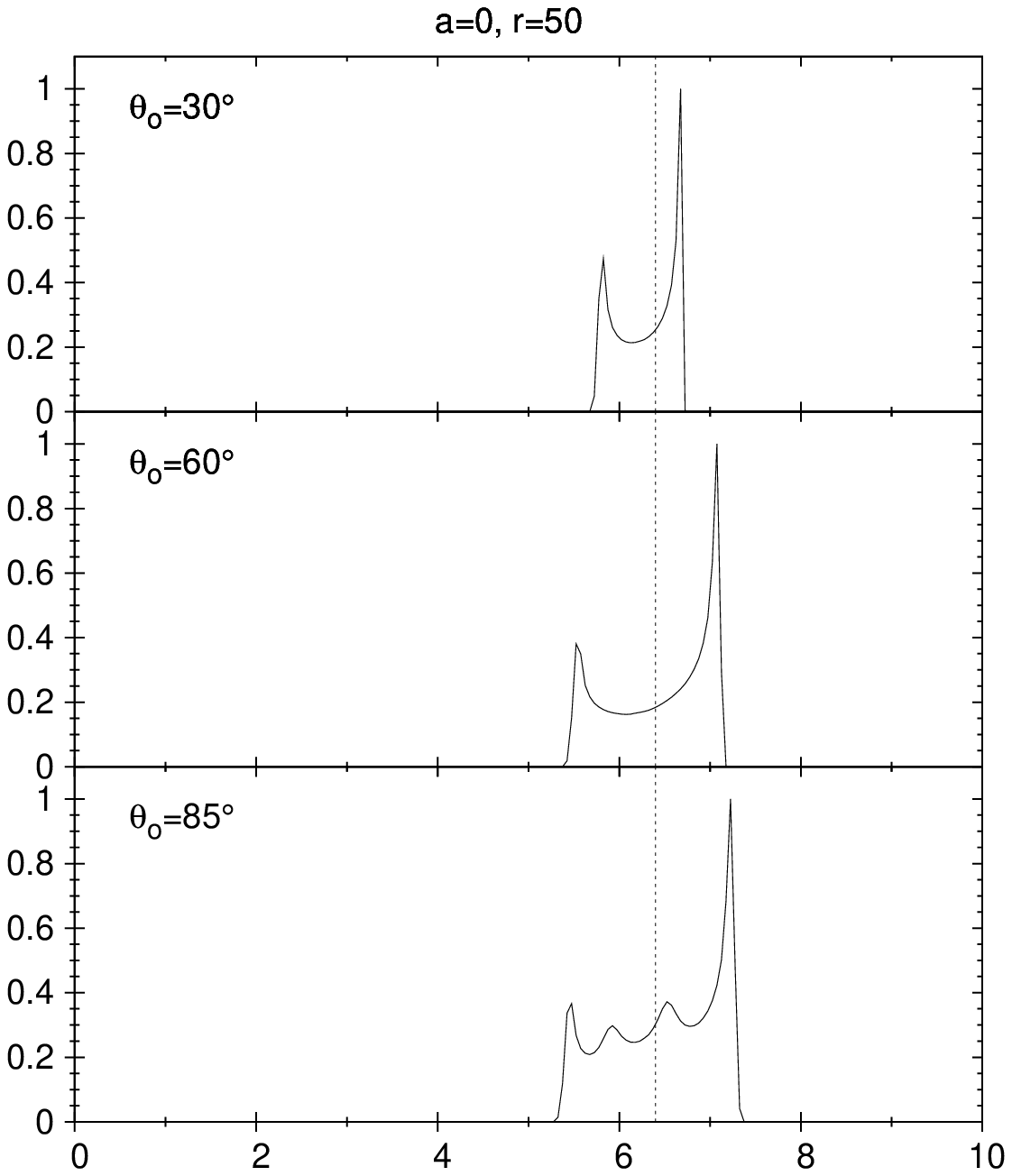}
\caption{Time--averaged synthetic spectra in terms of photon 
flux (in arbitrary units) versus energy (in keV). These 
profiles represent the mean, background-subtracted spectra
of the Fe K$\alpha$ iron-line originating from spots at
different radii. Top panels correspond to $r=r_{\rm{}ms}$ and
$a=0$ (left), $a=0.9$ (middle), and $a=1$ (right). In bottom panels 
we fix $a=0$ and choose $r=10$, $20$, and $50$, respectively
(other values of $a$ give very similar profiles).
Three consecutively increasing values of observer 
inclination $\theta_{\rm{}o}$ are shown, as indicated inside
the frames.}
\label{profiles}
\end{figure*}

In many cases, and especially for small radii and intermediate to large
inclination angles, the line emission comes from a relatively minor fraction
of the orbit. This implies in practice that for observations with a
{\it{}moderate signal-to-noise ratios, only a narrow blue horn can be visible, 
and only for a small part of the orbit.} 
These large and rapid changes of the line shape get averaged 
when integrating over the entire orbit, and so an important 
piece of information is missing in the mean spectra.
The line profiles integrated over the whole revolution 
are shown in Figure~\ref{profiles}. Effectively, the mean
profile of a spot is identical
to the profile of an annulus whose radius is equal to the
distance of the spot centre and the width is equal to the spot
size.

\section{Observational evidence for narrow, relativistic features}
Recently, interesting narrow emission features have been discovered in 
several AGNs. They occur in X-ray spectra, mostly (but not exclusively) 
at energies lower than $6.4$~keV, which is the intrinsic energy of the 
neutral iron K$\alpha$ fluorescent line in a frame co-moving with the 
emitting medium. The sources are: NGC~3516 (Turner et
al. 2002), ESO~198-G024 (Guainazzi 2003), NGC~7314 (Yaqoob et al. 2003),
and Mrk~766 (Turner et al. 2004). Possible interpretations include lines
from spallation products of iron (Skibo 1997), and shifted iron lines. In
the spallation model, the strongest expected line (after the iron line)
is the Cr~K$\alpha$ line at $5.4$~keV, the second strongest being the 
Mn~K$\alpha$ line at $5.9$~keV (Skibo 1997). Some of the observed features do
not match these energies (see below), and hence an energy shift is required.
However, a $6.4$~keV line is always present, suggesting that the {\it{}primary
line is not shifted}. We therefore consider in most cases the 
spallation model an unlikely explanation.

Shift of the iron line may be due either to fast moving ejecta, or to
orbiting material. Quality of the present data is not good enough to
discriminate between these two options. In the case of the orbiting spot model
presented in this paper, narrow features should occur in spectra because 
the two horns are often sharp. The blue horn is expected to be brighter 
than the red one (see Fig.~\ref{profiles}). For
insufficient signal-to-noise observations, only the blue horn may be
actually visible, the
remaining of the profile getting easily confused with continuum. It
is worth noting, however, that this argument does not apply to the extreme case 
of $a\rightarrow1$,
$r\,\rightarrow\,r_{\rm{}ms}$, where the profile is very broad and with no sharp
peak at all. Therefore, the observed features cannot arise from a spot orbiting
in the last stable orbit of a maximally rotating black hole.

In the following we will recapitulate the observational evidence for these features,
and derive the
system parameters in the framework of the orbiting spot scenario. 
For all these sources past claims of the presence of `classical' relativistic
lines do exist, although only for NGC~3516 the evidence seems to be robust. 
The equivalent widths (EW) of the narrow features are typically of a 
few tens of eV. Even allowing for the fact that only the blue horn is 
probably observed, this implies 
that the spot itself contributes just a moderate fraction of the total 
X-ray luminosity. In our estimate we assumed an isotropically illuminating 
flare (for definiteness of the model). It is worth noting that non-isotropic
illumination in the local frame co-moving with the flare is also possible,
and it would influence the predicted profiles especially if they originate
at small distance from the black hole. A relatively steady contribution 
to the signal is therefore 
required. Detailed temporal analysis of spectral features simultaneously 
with the continuum fluxes would be necessary in order to accurately test 
the scheme. Such analysis should also provide further constraints on
the angular distribution of illuminating spots. This is, however, beyond
the capabilities of present instruments, and therefore in the following we 
limit ourselves only to the discussion of spectral features. 

The comparison between the model and the observation will be made assuming 
that the emitted line has the rest-energy of $6.4$~keV, as measured in the 
local reference frame attached to the emitting matter. It means that
iron is not ionized more than Fe {\sc xvi}, a plausible 
assumption under usual conditions in accretion discs in AGNs. However, 
significant ionization of iron cannot be ruled out completely, 
especially below a flare, in which case
the rest-energy of the line could be larger (up to $6.97$~keV
for H--like iron). If this is the case, the energy shift must be also larger 
in order to achieve the same observed energy of the spectral feature, 
and this implies either a smaller radius or a smaller inclination angle 
with respect to the neutral line. Therefore, the values we will obtain for 
these two parameters should be considered as upper limits. 
This conclusion is rather straightforward but worth emphasizing: 
if the source inclination is estimated independently, then the adopted method 
can provide the {\it upper value to the radius where the line
emission originates}. Once the spectral resolution and sensitivity of 
the available instruments are improved, subtle differences will be 
exploited and the ionization state will be determined more 
precisely. For instance, the neutral fluorescent line and the H--like 
resonant lines are actually doublets, with different energy separation, 
which is not true for the resonant He--like line, not mentioning the 
shape for the Compton Shoulder and the ratio between the K$\beta$
and K$\alpha$ lines, which are dependent on ionization. 
There is thus some prospect to distinguish these cases in future.

It is worth noting that the expected narrow-line variability should be
associated with fluctuations of the X-ray continuum, because both spectral
components have common origin in this model. Indeed, power spectral
analysis of accreting black-hole sources shows clear evidence for such variability
on vastly different time-scales (e.g. Markowitz et al. 2003 and references 
therein). Recently, Collin et al. (2003) examined theoretical
spectra of the flare model, which is basically a generalization of the
lamp-post scheme to the case of a large number of independent primary sources 
occurring off-axis, just above the disc plane. It will be interesting to explore
in detail relations between continuum variability and the properties
of spectral features. This is however beyond the scope of this paper, and it is
deferred to a future work. We just note here that for two sources from our sample,
no variations of the feature during the observations are apparent, which implies
that neither variations in the continuum are expected. For the other two sources
there is indeed evidence for variations in the feature flux, and so 
for them the relation with continuum variations will be briefly discussed 
in relevant sections below.

\subsection{NGC 3516}
NGC~3516 is one of the best studied Seyfert 1 galaxies, and one of the 
best cases of
`classical', broad iron line from a relativistic disc (ASCA: 
Nandra et al. 1999; {\it{}XMM--Newton}: 
Turner et al. 2002). Evidence for a (possibly gravitationally)
redshifted iron absorption line in the ASCA spectrum was also reported by 
Nandra et al. (1999). 

NGC~3516 was observed twice by {\it{}XMM--Newton}, on April 2001 and
November 2001; both observations were partly overlapping with {\it{}Chandra}
ones. The November 2001 observations were published by Turner et al. (2002). 
The {\it{}Chandra}/HETG and {\it{}XMM--Newton} spectra exhibit five narrow lines.
One line is at $6.4$~keV (possibly originating in distant matter), two lines are seen 
bluewards (at $6.53$ and $6.84$--$6.97$~keV, the latter feature being variable and
recorded by {\it{}XMM--Newton} only), and two lines are redwards 
($5.57$ and $6.22$~keV, the former detected by {\it{}Chandra} only) of the rest
frame iron line energy. Turner et al. (2002) interpret these features as
the two horns of annular emission from $r=35$ and $r=175$, assuming
inclination angle of $\theta_{\rm{}o}=38^{\circ}$ (Wu \& Han 2001; this is actually
the angle estimated for the broad-line region (BLR) clouds distribution; 
the statistical error the authors quote is $\pm8^{\circ}$). Note, however, that
the black hole mass in NGC~3516 is estimated to be $\mbh=2.3\times10^7M_{\odot}$ 
(based on velocity dispersion; Wu \& Han 2001). This implies an orbital 
period of $T_{\rm{}orb}=150$~ks for $r=35$, and $T_{\rm{}orb}=1650$~ks 
for $r=175$. The former value of $T_{\rm{}orb}$ is comparable while the latter 
is significantly longer than the total {\it{}XMM--Newton} and {\it{}Chandra}
observing time (about $180$~ks). Both red and blue features have been detected in
the $75$~ks {\it{}Chandra} observation alone. Therefore, the annular emission must 
be steady, and not simply a consequence of integration over time of an orbiting 
spot signal. It is very hard to imagine a physical situation in which this 
can occur. Alternatively, the red and blue features may
be independent one another, the former being the blue horns from spots
near the black hole, the latter being emitted at large radii (time 
variation of the line centroid for one of the blue features is indeed 
suggestive of a spot orbiting with a period comparable or
longer than the exposure time).

We analysed the April 2001 {\it{}XMM--Newton} and {\it{}Chandra} observations (see
also Bianchi et al. 2004). We divided the $\sim$74 ks long {\it{}XMM--Newton}
exposure into three about equally spaced time intervals, about 17 ks each after
removing periods of high background. In all three time
intervals a narrow feature is observed at energy
$6.01\pm0.04$, $6.11\pm0.08$ and $6.04\pm0.11$~keV, respectively
(corresponding upper limits for $\sigma$ are uncertain; we obtained 
values of $\sigma=0.17$, $0.25$ and $0.88$~keV, respectively). 
The equivalent widths of these lines are ${\rm{}EW}=40$, $39$ and $21$~eV. 
The centroid energy of the feature is consistent with being constant 
as well as the flux (even if in the third interval the detection is marginal: 
the confidence levels at which the feature is detected are
$99.96$\%, $99.83$\% and $88.15$\%, respectively, according to the F-test). 
Indeed, summing the three time intervals together yields an energy of 
$6.08\pm0.03$ (the feature is significant at the $99.99$\% confidence level). 
We will call it the `red' feature. In all time intervals a narrow iron line 
at $6.4$~keV is also found which may originate in distant matter, while no 
evidence for the features detected by Turner et al. (2002) in the 
November 2001 observations is present. In the $73$~ks {\it{}Chandra} 
observation, instead, there is no evidence for features other than 
the $6.4$~keV line. The upper limit to the flux of a narrow line at
$6.1$~keV is, however, consistent with the flux measured in the 
{\it{}XMM--Newton} observation.

In our approach, the constancy in energy of the feature
suggests that the orbital period of the emitting annulus is
lower than the exposure time of a single interval, so that we are
averaging the profile over one or more orbits. Dividing further
in time the observation is not possible, as the signal-to noise 
ratio becomes too small. This feature could be the blue horn of a 
$r\sim6$ annulus profile (the rest of the line profile being 
too faint to be detectable). Corresponding inclination angle 
is slightly larger than $\theta_{\rm{}o}=30^{\circ}$ (cp. Fig.~\ref{profiles}). 
Indeed, a fit with a relativistic disc model ({\sc{}diskline} in XSPEC), 
and with inner and outer radii fixed to $r=6$ and $7$, respectively, 
comes out as good as the fit with a simple Gaussian line. The best fit
inclination angle is $\theta_{\rm{}o}=31^{\circ}$. In this case, the orbital period
is about $10$~ks, and for each time interval we would be averaging
over almost two orbits.

Alternatively, the `red' feature may be the red horn of a line profile
in which the $6.4$~keV line is the `blue' horn. Fitting the spectrum of
the whole observation with the {\sc diskline} model gives inclination
$\theta_{\rm{}o}=24^{\circ}$ and inner and outer radii of $r=12$ and $15.6$, 
respectively. At these radii the orbital period is about $T_{\rm{}orb}=30$~ks, i.e.
more than the observing time for each time interval, and so an orbiting spot cannot
provide the observed profile and we have to resort to the rather unpalatable steady
annular emission. Moreover, the statistical quality of the fit is poorer,
than with a double Gaussian (the latter fit giving an improvement
at the $99.95$\% confidence level), and the energy of the iron line, both
in {\it{}XMM--Newton} and {\it{}Chandra} (Turner et al. 2002) is suspiciously
close to the rest energy value. We therefore consider more likely that
the $6.4$~keV line originates in distant matter not affected by 
relativistic effects (such as the Broad Line Region or the
`torus'), and that the red feature is therefore the blue horn of a 
$r\sim6$ orbiting spot.

\subsection{ESO 198-G024}
The Seyfert 1 galaxy ESO 198-G024 has been observed by all major 
X-ray satellites. Behaviour of its $6.4$~keV iron line is rather puzzling:
it was found to be narrow in the ASCA spectrum and 
broad in the first {\it{}XMM--Newton} observation (Guainazzi 2003). 

The source was in fact observed by {\it{}XMM--Newton} twice: (i)~on December 1st,
2000 (Guainazzi 2003) and (ii)~on January 24th, 2001 (Porquet et al. 2003;
Bianchi et al. 2004). In the first observation, lasting about $9$~ks, a
narrow feature at $5.7^{+0.07}_{-0.12}$~keV was also detected with a confidence
level of $96.3$\% (${\rm{}EW}\sim70$~eV, but with $\sigma$ poorly constrained; 
the upper limit was found $0.28$~keV). Even if the statistical significance is 
rather marginal, it is tempting to liken this feature
with those reported in NGC~3516. However, we remind that 
the possibility cannot be ruled out that this feature is the red horn
of a `classical' double-peaked relativistic profile.
Notice that $6.4$~keV line with $\sigma=140^{+120}_{-70}$~eV was also
detected, which in this hypothesis would be interpreted as the blue horn
of the broad relativistic line; see discussion in Guainazzi 2003.

\begin{figure}
\epsfig{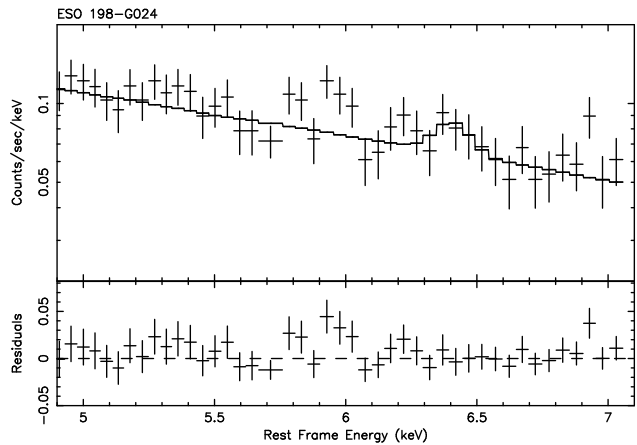}
\caption{Best fit spectrum and residuals for the second half of the second 
{\it{}XMM--Newton} observation of ESO 198-G024. Energy is given with respect
to the rest frame of the galaxy. The adopted model is a plain
power-law plus a Gaussian line to reproduce the $6.4$~keV line. 
Note the emission feature in residuals at $\sim5.96$~keV.}
\label{eso198}
\end{figure}

In the second observation, a narrow ($\sigma <0.1$~keV) feature is
observed at $5.96\pm0.05$~keV (again besides a $6.4$~keV line whose width
is loosely constrained, $\sigma<160$~eV; Bianchi et al. 2004), but only in 
the second half of the $\sim20$~ks observation (see Fig.~\ref{eso198}). 
The feature is significant at $98.3$\% confidence level, and the EW is 
about $65$~eV. The statistical significance of this feature is not very 
large, but its finding reinforces that of the previous observation.

One can assume that the $6.4$~keV line is of different origin, possibly 
arising in the disc or in BLR. (This interpretation however
seems to be incompatible with the line width in the first
observation, which is at $90$\% confidence level inconsistent with the
H$\beta$ line width of $6400$~km/s FWHM; Winkler 1992.) The other feature
detected in both observations could be the blue horn of an annulus with
$r\sim6$, seen at low inclination (see Fig.~\ref{orbits_1} and 
Fig.~\ref{profiles}). In that case the rest of the profile
would be below detectability. Furthermore, assuming (as an order of
magnitude estimate) that the visibility of the profile lasts for about a
quarter of the orbit (which corresponds to roughly $10$~ks), 
the resulting black hole mass is $\mbh\sim10^8M_{\odot}$.
However, we stress again that, given the poor statistics, the above
estimate should be taken more as a provisional hypothesis than as a 
reliable measurement.

A  brief mention on the continuum variability is appropriate here. 
The line equivalent width for an isotropic illumination is about 150 eV
(e.g. Matt et al. 1991). Given the observed value of EW, this means 
that the illuminating flare would account for about half the flux, 
and a $\sim$50\% continuum variability would have been expected to 
accompany the feature variability, which instead is
not observed. While stressing that the detection of the feature variability
must be considered as tentative, a possible situation is that the illumination
is very anisotropic and directed preferentially towards the disc 
(see e.g. Ghisellini et al. 1991).

\subsection{NGC 7314}
The Seyfert 1 galaxy NGC~7314 has a highly variable, broad iron line 
as observed by ASCA (Yaqoob et al. 1996). The source was then observed by the 
HETG onboard {\it{}Chandra}, and a narrow ($\sigma<30$~eV) feature 
at $5.84$~keV (${\rm{}EW}=20$~eV) was discovered (Yaqoob et al. 2003) 
in the low state,
after dividing the observation in two parts according to the count rate,
even if with a moderate significance ($\sim2\sigma$) only. As already
noted by these authors, the narrow feature may be the blue horn from an annulus
with $r\sim6$ at small inclination angle (see Fig.~\ref{orbits_1} and 
Fig.~\ref{profiles}). Another feature was also detected at $6.61$~keV 
(${\rm{}EW}=30~eV$); hence a viable alternative is that the latter feature 
represents the blue horn of emission from an annulus at $r\sim50$
and low inclination angle, with the $5.84$~keV being the red one, 
(see Fig.~\ref{profiles}). The mass of the black hole is estimated to 
be $\mbh\sim5\times10^6M_{\odot}$ (Padovani \& Rafanelli 1988),
implying an orbital period of $T_{\rm{}orb}=2.3$~ks at $r=6$, and
$T_{\rm{}orb}=55$~ks at $r=50$. The {\it{}Chandra} observation is 
$\sim100$~ks long, and so both explanations are viable as far as 
orbital time scales are concerned. To discriminate between the two
solutions, we re-analysed the {\it{}Chandra}
observation, dividing it in three time intervals of equal duration
about $30$~ks. In each time interval the feature is barely visible,
as expected (given the marginal detection) if the line flux is constant. 
This result therefore favours a small value of the orbital radius. 
However, given the limited quality of the data, no definitive conclusions 
can be drawn in this respect.

\subsection{Mrk 766}
Mrk~766 is one of the best studied narrow line Seyfert~1 galaxies. 
The presence of a broad, relativistic iron line in the spectrum
of this source is, however, still an open issue. Page et al. (2001) 
found evidence for a relativistic line in the first {\it{}XMM--Newton} 
observation. On the other hand, subsequent reanalysis of the same data 
set, as well as of the second {\it{}XMM--Newton} observation
(Pounds et al. 2003) did not provide evidence for such a line.

Turner et al. (2004) reported on the presence of a narrow feature 
at $5.6$~keV during the first $100$~ks of the second {\it{}XMM--Newton} 
observation, when the source was in a high state, and at $5.75$~keV 
in the last $30$~ks, after a sudden drop in flux. The EW of the
$5.6$~keV line is $\sim15$~eV, that of the $5.75$ about 4 times higher.
Turner et al. preferred explanation is in terms of a decelerating
ejected blob. Let us instead discuss the system parameters in the
orbiting spot hypothesis, even if the line and continuum variability
are anticorrelated, contrary to what is expected in our model. Clearly, if
the model is correct, continuum variability and illumination of the matter
must be very complex.
 
Assuming again that the observed narrow feature
corresponds to the blue horn, the emission must come from small radii to
account for the significant redshift. The mass of the black hole in this
source is estimated to be $10^7M_{\odot}$ (Wandel 2002),
corresponding to an orbital period of about $4.6$~ks for $r=6$, much less
than observing times for both states. Hence we can assume that the line
profile is averaged over several orbits. One possibility is that the
inclination angle is very low, in which case we can neglect Doppler shifts.
We find that the radius corresponding to the redshift-factor $g$
of a spectral line is equal to $2/(1-g^2)$ (in the Schwarzschild limit). 
The emission radius would then have been moved from about $r=8.5$ in 
the first part of the observation to $r=10.4$ in the second part. 
Increasing the inclination angle decreases the radius
corresponding to a given $g$ in the blue horn because of Doppler
blueshift. For $\theta_{\rm{}o}=30^{\circ}$ (see Fig.~\ref{profiles}) 
the radius is
already lower than $6$, requiring a spinning black hole. 
Of course, at least part of the line shift may alternatively
be due to ionization of the matter, which could also help explaining the larger
equivalent width (e.g. Matt et al. 1996). Clearly, high
throughput observations (capable of detecting line features in short 
exposure times) are necessary to break the degeneracy between $r$ and 
$\theta_{\rm{}o}$.

\section{Conclusions}
We discussed the possibility that the narrow features in the 
$5$--$6$~keV range, recently discovered in a few AGNs and 
usually interpreted as
redshifted iron lines, could be due to illumination by localized
orbiting spots just above the accretion disc. If this is indeed the
case, these features may provide a powerful and direct way to measure
the black hole mass in active galactic nuclei. To this aim, it is
necessary to follow the line emission along the orbit. 
The orbital radius (in units of $r_{\rm{}g}$) and the disc inclination 
can be inferred from the variations of the line flux and centroid energy.
Furthermore, $\mbh$ can be estimated by comparing the measured orbital 
period with the value expected for the derived radius. 
As shown in the previous section, present-day X-ray
instruments do not have enough collecting area to perform this task
accurately. This capability should be achieved by the planned 
high-performance X-ray missions such as
{\it{}Constellation--X} and {\it{}Xeus}.

\section*{Acknowledgements}
We thank A.~Martocchia and the anonymous referee
for useful comments. This paper is based partly on observations 
obtained with {\it{}XMM--Newton}, an ESA science mission with 
instruments and contributions directly funded by ESA Member
States and the USA (NASA). VK and MD acknowledge support from 
grants GACR 205/03/0902 and 202/02/0735.
SB and GM acknowledge financial support from Italian ASI and MIUR. 

{}
\label{lastpage}
\end{document}